\documentclass{article}
\usepackage[main, final]{neurips_2026}  %
\usepackage{times}
\usepackage{graphicx}
\usepackage{amsmath,amssymb,amsthm}
\usepackage{booktabs}
\usepackage{multirow}
\usepackage{subfig}
\usepackage{algorithm}
\usepackage{algorithmic}
\usepackage{xcolor}
\usepackage{enumitem}
\usepackage{colortbl}
\usepackage{pifont}
\usepackage[most]{tcolorbox}
\usepackage{hyperref}

\newcommand{\ours}{\textsc{Harp}}

\newtheorem{lemma}{Lemma}
\newtheorem{remark}{Remark}

\definecolor{promptHead}{HTML}{1F6B3A}
\definecolor{promptBody}{HTML}{F4F7F2}
\definecolor{caseHead}{HTML}{1F3A6B}
\definecolor{caseBody}{HTML}{F2F5FA}

\newtcolorbox{promptbox}[1]{%
  enhanced, breakable,
  colback=promptBody, colframe=promptHead,
  coltitle=white, fonttitle=\bfseries\small,
  title={#1},
  left=6pt, right=6pt, top=4pt, bottom=4pt,
  boxrule=0.6pt, arc=2pt,
  fontupper=\small\ttfamily,
  before skip=6pt, after skip=6pt,
}

\newtcolorbox{casebox}[1]{%
  enhanced, breakable,
  colback=caseBody, colframe=caseHead,
  coltitle=white, fonttitle=\bfseries\small,
  title={#1},
  left=6pt, right=6pt, top=4pt, bottom=4pt,
  boxrule=0.6pt, arc=2pt,
  fontupper=\small,
  before skip=6pt, after skip=6pt,
}

\title{\textsc{Harp}: Training-Free Dual-Profile Agentic Communication for LLM-Based Recommendation}

\author{%
  \textbf{Zhen Tao$^{1,2}$, Xun Zhou$^{1,3,*}$, Xuhui Chen$^{1}$, Ziyue Qiao$^{2}$, Qingqiang Sun$^{2,*}$} \\
  $^{1}$Harbin Institute of Technology, Shenzhen \\
  $^{2}$Great Bay University, \quad $^{3}$Shenzhen Loop Area Institute \\
  \texttt{25b951132@stu.hit.edu.cn, zhouxun2023@hit.edu.cn,} \\
  \texttt{25b951013@stu.hit.edu.cn, ziyuejoe@gmail.com, qqsun@gbu.edu.cn}
}

\begin{document}
\maketitle
{\let\thefootnote\relax\footnotetext{$^{*}$Corresponding author.}}

\begin{abstract}
In LLM-based recommendation, \emph{memory} effectively plays the role of a user-level agent, typically built on top of a memory-augmented backbone.  Yet under this paradigm users and items remain asymmetric participants: only the user is given a structured profile; that profile is consumed inside a bandwidth-limited reranker prompt; and it is frozen after warmup.  Starting from this observation, we recast LLM-based recommendation as an \emph{agentic-communication} process in four stages: \textsc{Build} a self-description, \textsc{Transmit} it along the collaborative graph, \textsc{Understand} the other side, and \textsc{Sync} it from the next interaction.  Analysing the dominant memory-augmented paradigm through this formulation, we identify three open communication challenges that current pipelines leave unaddressed: the item side never produces a comparable self-description (one-sided \textsc{Build}); profile-level signals are routed only into a bandwidth-limited reranker prompt rather than into the retrieval stage where they would not compete for context tokens (prompt-trapped \textsc{Understand}); and the user's description is frozen after warmup with no principled update rule (absent \textsc{Sync}). 
We instantiate the missing stages as \textbf{\ours{}} (\emph{\underline{H}omophily-aware \underline{A}gentic \underline{R}ecommendation via dual \underline{P}rofiles}), a training-free, three-operator addition to a frozen LLM backbone: a rule-extracted symmetric item profile (zero extra LLM calls), a channel-decomposable profile-level homophily $\mathcal{H}(\phi_u,\phi_i)$ consumed at retrieval time rather than inside the reranker prompt, and a single-pass reflective update realising \emph{in-context} profile evolution without any gradient step.  On four public benchmarks (Yelp, Books, MovieTV, Goodreads), \ours{} lifts Hit@1 over the strongest published baseline by $6.9\%$--$24.0\%$, with family-wise Holm-corrected $p<10^{-9}$ on every dataset.  Ablations and a per-case attribution study show that the operators contribute complementary gains, with the retrieval-side gate as the main source; we also report the trade-off introduced by the reflective pass.  Our contribution is a unified formulation of agentic communication for LLM-based recommendation, together with a training-free three-operator instantiation on top of it. 
\end{abstract}

\section{Introduction}
\label{sec:intro}

\begin{figure*}[t]
\centering
\includegraphics[width=0.98\textwidth]{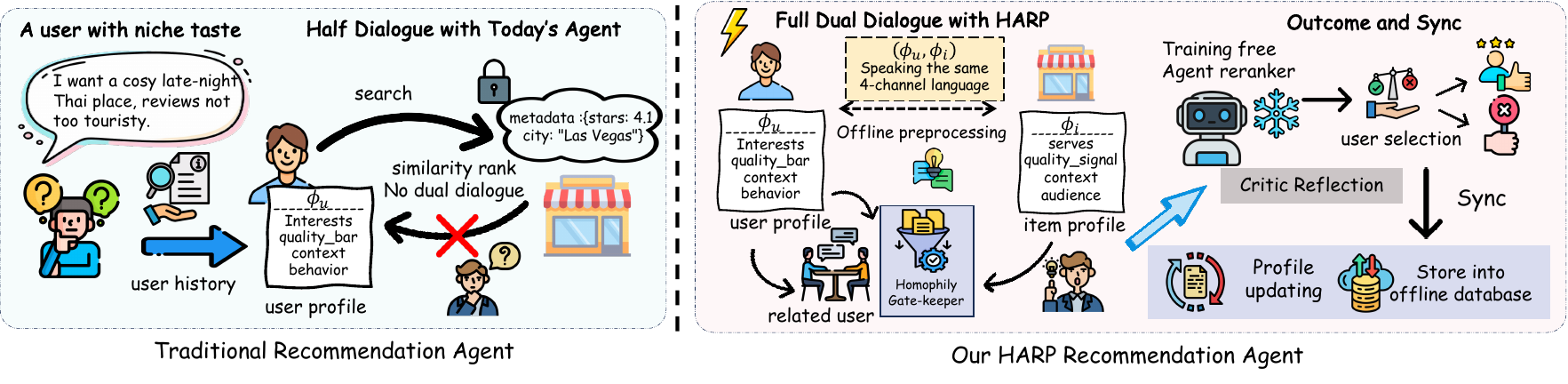}
\vspace{-1mm}
\caption{\ours{} at a glance. A user with a niche query in a half-dialogue session, compared with the full dialogue restored by \ours{}. HARP integrates three training-free innovations: Dual Profile (\S\ref{sec:profile}), Homophily Gate (\S\ref{sec:communication}), and Reflection (\S\ref{sec:evolution}).}
\label{fig:motivation}
\end{figure*}

Large language models (LLMs) are becoming the central reasoning engine of next-generation recommender systems~\cite{wu2024survey,geng2022recommendation,liu2023chatgpt,zhang2025towards}.  In the most representative \emph{agentic recommendation} paradigm~\cite{zhang2024agentcf,xu2025iagent,chen2026memrec,tang2026interactive}, a user is modelled as an LLM agent with persistent memory, and personalised recommendations are produced by routing collaborative-neighbour memories into a reranker prompt.  Yet this paradigm provides only a partial answer: the user side is equipped with memory while the item side has no structured profile aligned with the user's; the user profile is consumed solely inside a bandwidth-limited reranker prompt; and the profile built at warmup is not revised from recommendation outcomes afterwards.  We recast LLM-based recommendation as a four-stage agentic-communication process between user and item agents (\textsc{Build}--\textsc{Transmit}--\textsc{Understand}--\textsc{Sync}), and show that the dominant paradigm implements only a partial version of it, leaving three communication challenges open (Figure~\ref{fig:motivation}).

\paragraph{The four-stage framing and a motivating observation.}
The stages \textsc{Build}--\textsc{Transmit}--\textsc{Understand}--\textsc{Sync} borrow directly from heterogeneous message passing on graphs~\cite{hu2020heterogeneous, schlichtkrull2018modeling} and from the classical study of homophily on social networks~\cite{mcpherson2001birds}.  The decomposition plays a \emph{dual} role: as an \emph{analytical tool} that asks which of the four stages a given pipeline implements and how well, and as a \emph{generative scaffold} that prescribes the operator to add once a missing stage is identified.  The four-stage view is thus an analytical framework for diagnosing a pipeline, not a model in itself.  Recommender systems have passed through three architectural generations (collaborative filtering~\cite{koren2009matrix, sarwar2001item}, graph neural networks~\cite{he2020lightgcn, velivckovic2017graph}, and the recent agentic paradigm~\cite{wu2024survey}), yet the collaborative signal pervading all three is indexed by interaction \emph{overlap} rather than description-level compatibility, hence size-blind: whenever two agents' description sets differ sharply in breadth, raw interaction-overlap similarity inflates the match for reasons of degree rather than relevance, and the pre-rerank candidate list inherits that inflation.  An agentic communication pipeline therefore needs a profile-level, \emph{asymmetric} coverage measure read from the query side's vantage, which is the role the channel-decomposable homophily $\mathcal{H}$ of \S\ref{sec:profile} plays once the item side is also allowed to speak.

\paragraph{Three communication challenges and three operators.}
Applied to the memory-augmented family on four public benchmarks (Yelp, Amazon Books, MovieTV, Goodreads), this analysis identifies three open challenges where the communication falls silent.  \textbf{(Challenge~1) Asymmetric \textsc{Build}.} The user side carries a multi-field profile while the item side has no profile in the same schema.  \textbf{(Challenge~2) Bandwidth-trapped \textsc{Understand}.} Concatenating the user profile into the reranker does not move ranking accuracy: by the time the reranker is called, its context is fixed and the context budget is dominated by the memory block; \textsc{Understand} is the stage where profile-level signal must enter \emph{before} the reranker is called.  \textbf{(Challenge~3) Absent \textsc{Sync}.} Profiles built at warmup are not calibrated against the feedback that warmup itself produces.

Each challenge is addressed by a primitive already validated elsewhere, composed without retraining.  \textbf{Against Challenge 1}, a structured-JSON user profile, in the spirit of agent-maintained profiles in prior agentic recommenders~\cite{zhang2024agentcf}, is used on the user side, and a schema-aligned item profile is introduced via rule extraction at no extra LLM cost.  \textbf{Against Challenge 2}, the collaborative retrieval score is re-weighted by a profile homophily that generalises the scalar edge-homophily ratio~\cite{zhu2020beyond} to a sum over semantic channels; homophily is used here as an analytical quantity, not as an optimisation target.  \textbf{Against Challenge 3}, a single Chain-of-Thought reflective pass~\cite{wei2022chain,shinn2023reflexion,madaan2023self} optionally guarded by a conservative critic is added, realising \emph{in-context evolution} of the user profile without any gradient update, a \emph{self-evolving}-agent primitive in the sense of the recent survey~\cite{gao2025survey}.  The composition is \textbf{\ours{}} (\emph{\underline{H}omophily-aware \underline{A}gentic \underline{R}ecommendation via dual \underline{P}rofiles}), a training-free addition to a frozen LLM backbone that operates entirely through the \emph{LLM-as-agent} interaction loop; its novelty lies in the systematic identification of these three primitives as the complementary response to the three challenges within the memory-augmented agentic family, each separable in ablation (\S\ref{sec:experiments}).

\paragraph{Our main contributions are as follows:}
\begin{itemize}[leftmargin=*, itemsep=1pt, topsep=2pt]
\item \textbf{Formulation.} An \emph{agentic-communication} formulation that models LLM-based recommendation as a four-stage process \textsc{Build}--\textsc{Transmit}--\textsc{Understand}--\textsc{Sync} and, used analytically, surfaces the three communication gaps that the dominant memory-augmented agentic paradigm leaves open.
\item \textbf{Method.} \ours{}, a three-operator training-free addition to a frozen LLM backbone: a rule-extracted symmetric item profile, a channel-decomposable homophily-gated retrieval, and a single-pass reflective update, each aligned with one of the three gaps.
\item \textbf{Design rationale.} Three design disciplines that predict where operator-level gains should appear: the channel-decomposability of profile-level homophily, the retrieval-vs-prompt design rationale, and the one-pass conservative update.
\item \textbf{Empirical evaluation.} On four public benchmarks, \ours{} achieves state-of-the-art on every metric, lifting Hit@1 by $6.9\%$--$24.0\%$ over the strongest published baseline; ablations and per-case traces further confirm the effectiveness and complementarity of each operator.
\end{itemize}

\section{Related Work}
\label{sec:related}

\paragraph{LLM-based agentic recommenders.}
The agentic paradigm represents users as LLM-managed stateful memories~\cite{zhang2024agentcf, xu2025iagent, tang2026interactive, huang2025recommender}, yielding a chronology from stateless vanilla LLM~\cite{liu2023chatgpt}, static per-user memory~\cite{huang2025recommender, xu2025iagent}, dynamic self-reflective memory~\cite{zhang2024agentcf, xu2025iagent, tang2026interactive}, to a dynamic memory graph managed by a lightweight LM~\cite{chen2026memrec}.  Within this line, RecMind~\cite{wang2024recmind} augments user agents with tool-calling for external knowledge, and RecAgent~\cite{wang2025user} simulates realistic user behaviour with LLM-driven decision loops.  Related work also improves modality representations for recommendation~\cite{tang2025r2mr}, uses LLMs to understand user text in conversational recommenders~\cite{yuan2026integrating}, and generates semantic item identifiers for cross-domain recommendation~\cite{hu2026ids}.  A parallel infrastructure thread provides pluggable agentic memory backends: A-MEM~\cite{xu2026mem} proposes structured agentic memory for LLM agents, H-Mem~\cite{ye2026hmem} introduces hybrid multi-dimensional memory for long-context conversations, and a recent task-aware retrieval thread~\cite{tao2026task} makes the retrieval step itself context-adaptive.  \ours{} differs by lifting \emph{both} sides to profile level and consuming the resulting homophily at retrieval time rather than inside a reranker prompt.

\paragraph{Homophily in networks and recommender systems.}
Homophily, the principle that similarity breeds connection~\cite{mcpherson2001birds, lazarsfeld1954friendship}, has been studied inside GNNs~\cite{zhu2020beyond, ma2021homophily, platonov2023critical, luan2023graph, zheng2026graph} where message passing degrades on heterophilous graphs, and graph contrastive and fairness-aware learning likewise exploit neighbourhood structure~\cite{sun2025single, sun2024interdependence, li2026fairness}; classical CF~\cite{koren2009matrix,sarwar2001item,he2017neural,he2020lightgcn,wang2019neural} implicitly relies on homophily at the interaction level.  Recent analyses sharpen the picture: \citet{platonov2023critical} challenge existing heterophily benchmarks, \citet{luan2023graph} characterise when homophily helps node classification, and \citet{zheng2026graph} survey heterophily-aware GNN designs.  These works, including the cross-type homophily of heterogeneous graphs~\cite{tao2025cross}, operate on \emph{interaction-level} or \emph{class-label} homophily; \ours{} differs by defining $\mathcal{H}$ at the \emph{profile} level: agent-authored, agent-mutable, and additively decomposable over four semantic channels (Remark~\ref{thm:decomp}).

\paragraph{Self-evolving agents and agentic RAG.}
Recent work shows LLM agents can self-improve via in-context rewrites~\cite{gao2025survey, zhang2026agentic, wang2023voyager, shinn2023reflexion, madaan2023self} or make retrieval decisions adaptively~\cite{singh2025agentic, asai2024self, jiang2023active}.  Voyager~\cite{wang2023voyager} accumulates reusable skills in Minecraft, ChatDev~\cite{qian2024chatdev} organises software tasks as multi-agent role-play dialogues, and Self-RAG~\cite{asai2024self} trains the model to decide when and what to retrieve.  In the agentic-context-engineering view~\cite{zhang2026agentic}, the agent's prompt itself becomes an evolvable artefact, a perspective \ours{} instantiates for recommendation profiles.  \ours{} differs by applying an acceptance check (Eq.~\ref{eq:accept_rule}) to profile evolution and routing the homophily signal to the retrieval stage rather than the generation stage.

\noindent Detailed discussion of all adjacent threads is deferred to Appendix~\ref{app:rw_extended}.

\section{The \ours{} Method}
\label{sec:methodology}

This section develops \ours{} in four steps: \S\ref{sec:overview} fixes the recommendation pipeline and notation; \S\ref{sec:profile}--\S\ref{sec:evolution} introduce the three operators that separate \ours{} from a one-sided agentic backbone; \S\ref{sec:failure_modes_ref} aligns each operator with the structural challenge its ablation re-introduces.  Full notation, complete pseudocode, complexity analysis and dataset-level extractor details are deferred to Appendix~\ref{app:notation}--\ref{app:item_extractor}.

\begin{figure*}[t]
\centering
\includegraphics[width=0.99\textwidth]{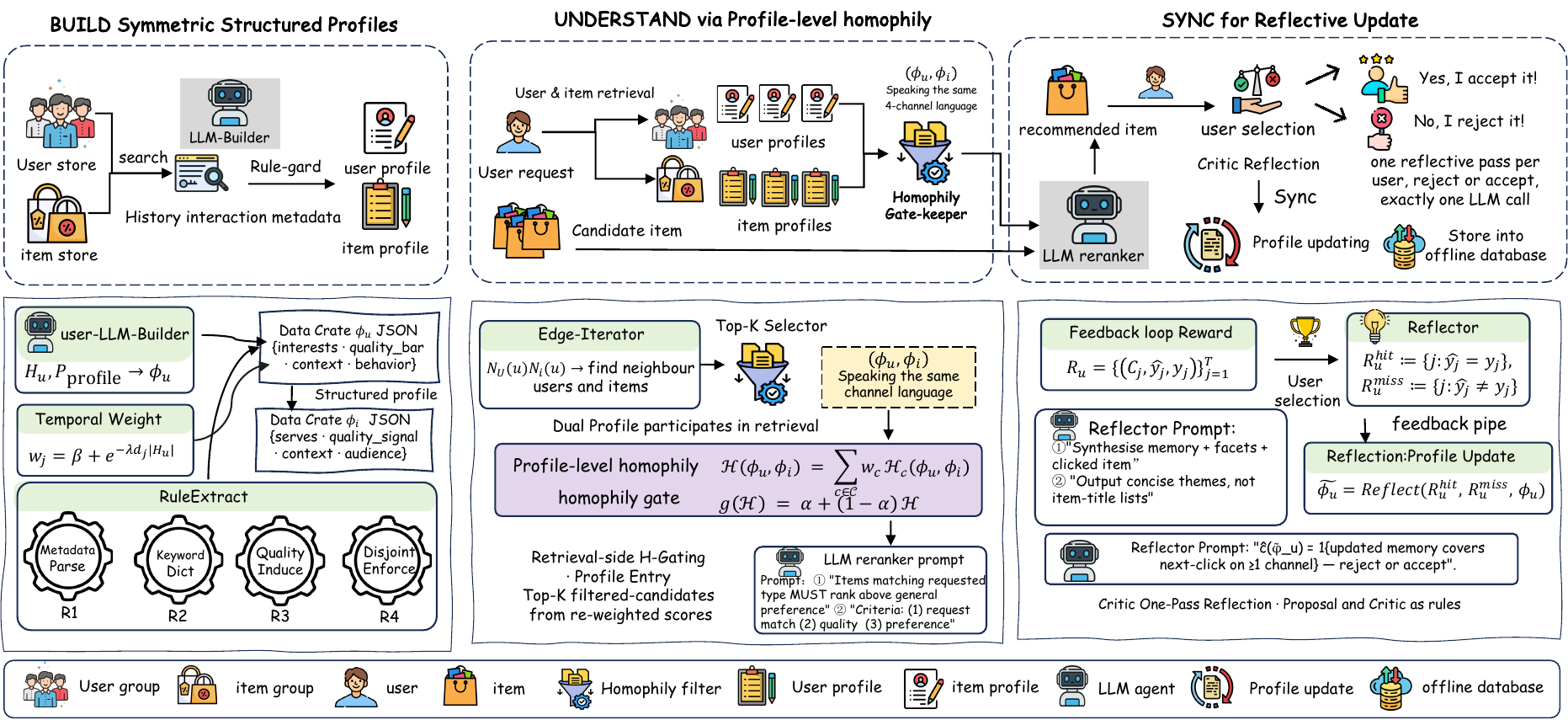}
\caption{\ours{} pipeline overview. \emph{Upper}: story strip illustrating the flow of a user's query across the entire pipeline. \emph{Lower}: block-exploded technical view with input/output annotations and equation references.}
\label{fig:framework}
\end{figure*}

\subsection{Problem Setup and Preliminaries}
\label{sec:overview}

\paragraph{Two-stage agentic recommendation and the core design.}
We follow the standard two-stage setup in current agentic recommenders~\cite{zhang2024agentcf,xu2025iagent,tang2026interactive}.  Let $\mathcal{V}_U,\mathcal{V}_I$ denote the user and item sets, $H_u\subseteq\mathcal{V}_I$ user $u$'s interaction history, $\mathcal{I}_u$ the evaluation instruction, and $\mathcal{N}_U(u),\mathcal{N}_I(u)$ the one-shot $k$-nearest user and item neighbour sets of $u$.  For user $u$, a retriever selects $k$ collaborative neighbours $N(u)$ (Eq.~\ref{eq:retrieval}); a frozen LLM reranker selects the top-1 prediction from the candidate list $C(u)$ given by the evaluation protocol (Eq.~\ref{eq:rerank}):
\begin{equation}
N(u) \;=\; R(u;\,H_u,\,\mathcal{N}_U(u)\cup\mathcal{N}_I(u)) \;=\; \mathrm{Top\text{-}k}_{\,n}\;\mathrm{BaseScore}(u,n),
\label{eq:retrieval}
\end{equation}
\begin{equation}
\hat y_u \;=\; \arg\max_{i\in C(u)}\; \mathrm{LLM}_{\mathrm{rerank}}(\mathcal{I}_u,\,M_{\mathrm{collab}},\,\phi_u,\,C(u)),
\label{eq:rerank}
\end{equation}
where $M_{\mathrm{collab}}$ is a memory block aggregated along the selected neighbours $N(u)$ and $\phi_u$ is an optional user profile.  On top of this shared backbone, \ours{} introduces a layer of \emph{channel-level understanding} \emph{above} memory: both user and item carry a structured profile whose four channel fields $\phi_v=(\phi_v^{(c)})_{c\in\mathcal{C}}$ are aligned over the channel set $\mathcal{C}=\{\mathrm{preference},\mathrm{quality},\mathrm{context},\mathrm{audience}\}$, whose per-channel fields are phrase sets disjoint by construction (R1--R4 in \S\ref{sec:profile}; Table~\ref{tab:channels}).  This aligned schema lets user--item matching decompose into separately measurable semantic channels, and turns \emph{profile-level homophily} (Eq.~\ref{eq:homophily_def}) into an explicit quantity used for retrieval-side gating and post-warmup reflective rewriting.  The prevailing memory-augmented family corresponds to the special case in which three restrictions are simultaneously imposed (\emph{absent item profile, profile injected only via prompt, profile never updated}); the three operators in \S\ref{sec:profile}--\S\ref{sec:evolution} lift these in turn.  Three operational constraints run through the section: \textbf{(C1)} the reranker prompt has finite token budget $B$, so any content in the $\phi_u$ slot trades against $M_{\mathrm{collab}}$; \textbf{(C2)} the backbone LLM is frozen; \textbf{(C3)} each user receives at most one accepted reflective-update trajectory (applied at end of warmup; the trajectory may itself consist of $T_{\mathrm{evo}}$ reflection calls at different history offsets, with $T_{\mathrm{evo}}=1$ in main experiments).  A full notation table is in Appendix~\ref{app:notation}.

\subsection{\textsc{Build} Operator: Symmetric Structured Profiles}
\label{sec:profile}

\paragraph{Two-sided builder.}
The user profile is produced by a single LLM call over $H_u=H_u^{\mathrm{recent}}\sqcup H_u^{\mathrm{medium}}\sqcup H_u^{\mathrm{distant}}$ under a fixed JSON-schema prompt $P_{\mathrm{profile}}$:
\begin{equation}
\phi_u \;=\; \mathrm{LLM}\bigl(H_u^{\mathrm{recent}}\,\|\,H_u^{\mathrm{medium}}\,\|\,H_u^{\mathrm{distant}}\,\|\,P_{\mathrm{profile}}\bigr),
\label{eq:user_profile}
\end{equation}
where the three bands are relative positions in the history, newest to oldest: recent interactions keep full detail (title, categories, rating, city, first review sentence), medium ones are compressed to category and city statistics, and distant ones to the domain boundaries they touch, with an Ebbinghaus-style temporal weight $w_j = \beta + \exp(-\lambda\, d_j\,|H_u|)$ per observation $j\in H_u$ ($d_j\in[0,1]$ normalised age, $\lambda>0$ forgetting rate, $\beta>0$ floor)\label{eq:forgetting}; the banding realises this weight under a fixed prompt budget while keeping long-range coverage.  The seven-field JSON schema is designed for this paper, following the idea of agent-maintained user profiles in prior agentic recommenders~\cite{zhang2024agentcf}; only the four \emph{channel} fields in Table~\ref{tab:channels} participate in the homophily computation (Eq.~\ref{eq:homophily_def}).  The item side is obtained via rule extraction, with zero LLM calls:
\begin{equation}
\phi_i \;=\; \mathrm{RuleExtract}(\mathrm{meta}_i,\,\mathrm{reviews}_i),
\label{eq:item_profile}
\end{equation}
targeting the same channel-aligned schema, which keeps item-side profiling feasible at $N_I\sim 10^4$--$10^5$.  $\mathrm{RuleExtract}$ is formalised as a \emph{deterministic, zero-LLM} shallow semantic parser, fully parametrised by a dataset-level ontology map $\mathcal{S}$ and domain phrase lexicon $\mathcal{D}$, executed in four rules R1--R4 (ontology projection, lexicon-guided encoding, quality-signal induction, channel-exclusivity projection) that route metadata and reviews into the four pairwise-disjoint channel fields; per-rule definitions, per-dataset lexicons, and coverage statistics are in Appendix~\ref{app:item_extractor}.

\begin{table}[h]
\centering
\small
\caption{Four-channel correspondence between user and item profiles.}
\label{tab:channels}
\begin{tabular}{lll}
\toprule
\textbf{Channel $c$} & \textbf{User field} $\phi_u^{(c)}$ & \textbf{Item field} $\phi_i^{(c)}$ \\
\midrule
preference & \texttt{interests}       & \texttt{serves} \\
quality    & \texttt{quality\_bar}    & \texttt{quality\_signal} \\
context    & \texttt{context}         & \texttt{context} \\
audience   & \texttt{behavior}        & \texttt{audience} \\
\bottomrule
\end{tabular}
\end{table}

\paragraph{Profile-level homophily as a similarity score.}
For the phrase-set channels \texttt{preference} and \texttt{audience}, the similarity between two phrase sets $A,B\subseteq\Sigma^*$ is
\begin{equation}
\mathrm{sem}(A,B) \;=\; \frac{\bigl|\{a\in A\,:\,\exists\,b\in B,\;\mathrm{substr}(a,b)\lor\mathrm{substr}(b,a)\}\bigr|}{|A|}\;\in\;[0,1],
\label{eq:soft_jaccard}
\end{equation}
with $\mathrm{sem}=0.5$ (neutral) when $A$ or $B$ is empty.  The \emph{asymmetric normalisation} is deliberate: fixing the denominator on the \emph{query side} ($A=\phi_u^{(c)}$) makes Eq.~\ref{eq:soft_jaccard} read ``how much of what $A$ claims is actually echoed by $B$''; the relevance perspective is anchored to the querying user, and a broader-vocabulary item cannot gain extra credit by mere size.  The \texttt{quality} and \texttt{context} channels, whose fields are not phrase sets, use channel-specific fallback implementations (Appendix~\ref{app:sem_detail}).  All three forms participate uniformly in Eq.~\ref{eq:homophily_def}, and profile-level homophily is defined as a channel-wise weighted average:
\begin{equation}
\mathcal{H}(\phi_u,\phi_i) \;=\; \sum_{c\in\mathcal{C}}\,w_c\,\mathcal{H}_c(\phi_u,\phi_i), \qquad \mathcal{H}_c(\phi_u,\phi_i) \;:=\; \mathrm{sem}\bigl(\phi_u^{(c)},\,\phi_i^{(c)}\bigr),
\label{eq:homophily_def}
\end{equation}
where $(w_c)_{c\in\mathcal{C}}$ are non-negative channel weights with $\sum_c w_c=1$; the \texttt{preference} channel usually takes half of the weight or more and the other three share the rest evenly; the weights are adjustable hyperparameters that are \emph{not} tuned against any downstream metric.  When profiles degenerate to a single class label, $\mathcal{H}$ reduces to the same-class indicator of an edge, whose average over all edges is the edge-homophily ratio of~\cite{zhu2020beyond}; $\mathcal{H}$ is thus a natural generalisation of the graph-homophily concept to the \emph{multi-channel, multi-granularity semantic} setting.  In \ours{}, $\mathcal{H}$ plays the dual role of \emph{diagnostic quantity} and \emph{retrieval-side prior}, not an end-to-end optimisation target.

\begin{remark}[Channel decomposability: algebraic basis for separable ablations]
\label{thm:decomp}
Pairwise disjointness of channels (guaranteed by R4) together with the linear aggregation of Eq.~\ref{eq:homophily_def} imply: replacing channel $c$'s phrase sets $(\phi_u^{(c)},\phi_i^{(c)})$ while holding other channels fixed changes $\mathcal{H}$ by exactly $w_c\,\Delta\mathcal{H}_c$, with no cross-channel coupling.  This equates ``per-channel ablation'' with ``additive perturbation of the aggregate $\mathcal{H}$'', so changes in $\mathcal{H}$ can be attributed additively to individual channels.  Full proof in Appendix~\ref{app:proof_decomp}.
\end{remark}

\subsection{\textsc{Understand} Operator: Homophily-Gated Retrieval}
\label{sec:communication}

\paragraph{Homophily-gated neighbour score.}
Classical collaborative retrieval propagates preference signals via \emph{behavioural similarity} on $\mathcal{N}_U(u)$, implicitly assuming ``behavioural neighbours are preference neighbours.''  This assumption fails in sparse, heterogeneous settings via \emph{cross-category misleading neighbours}.  We use the profile-level homophily $\mathcal{H}(\phi_u,\phi_i)$ from \S\ref{sec:profile} as a \emph{semantic prior} to re-weight the behavioural signal.  Introducing a monotone homophily gate $g(\mathcal{H}) = \alpha + (1-\alpha)\,\mathcal{H}$ ($\alpha\in[0,1]$ the homophily floor), we obtain the gated neighbour score
\begin{equation}
\mathrm{CommScore}(u,n) \;=\; \mathrm{BaseScore}(u,n)\cdot g\!\bigl(\mathcal{H}(\phi_u,\phi_{i_n})\bigr),
\label{eq:comm_score}
\end{equation}
where $i_n$ is an item neighbour of $u$ and $\alpha$ ensures that collaborative signal is not fully zeroed when $\mathcal{H}\to 0$.  The gated quantity is the \emph{cross-type} profile homophily $\mathcal{H}(\phi_u,\phi_{i_n})$ (cf.~\cite{tao2025cross}): when the channel set degenerates to a single class label, $\mathcal{H}$ is the same-class indicator of the edge, the per-edge quantity behind the edge-homophily ratio of~\cite{zhu2020beyond}; the multiplicative gate itself is our design.  $g$ is a \emph{monotone bounded} transform of $\mathcal{H}$, so order-preservation guarantees the gate does not distort the channel-induced ordering; at $\alpha=1$ or $\mathcal{H}\equiv 1$ it degenerates to a pure behavioural retriever.  We treat user--user edges ($\phi_u,\phi_n$ via symmetric phrase overlap on the \texttt{preference} channel) and user--item edges as two instances of the same \textsc{Understand} operator, with independent floors (main experiments: $\alpha_{ui}=0.4,\,\alpha_{uu}=0.3$) acting on disjoint neighbour subsets.

\paragraph{Where in the pipeline should the profile be consumed?}
A non-trivial design choice is whether profile signal should enter via \emph{retrieval-side candidate reshaping} or \emph{reranker-side context injection}.  The two are \emph{not} information-equivalent: the former changes \emph{which neighbours} inform the reranker, the latter changes the \emph{evidence} it relies on.  We argue the former is strictly preferable here for two reasons.  \emph{Budget} (induced by C1): the reranker's context token budget $B$ is hard; once $B$ is saturated, every token added to the $\phi_u$ slot displaces $M_{\mathrm{collab}}$, so prompt-side injection is effectively a \emph{zero-sum substitution} between ``profile evidence'' and ``collaborative memory''.  \emph{Mechanism}: under C2 (frozen LLM), the reranker cannot learn to balance the two sources, and its input is already fixed by the retrieved candidates; with misleading candidates, even the optimal profile context cannot reliably promote $y_u$.  Placing the profile at the retrieval side lets it act on the \emph{recoverable stage}.  We record this intuition formally as Remark~\ref{prop:budget} in Appendix~\ref{app:prop_budget}; Table~\ref{tab:ablation} in \S\ref{sec:experiments} provides the empirical comparison (the two routes are labelled \textbf{CH} and \textbf{RI} in the ablation).

\subsection{\textsc{Sync} Operator: Reflective Update}
\label{sec:evolution}

\paragraph{Warmup record and reflective proposal.}
Under C2 (frozen backbone), the model cannot absorb user feedback via gradient flow; but each user $u$'s interaction history $H_u$ naturally constitutes a \emph{self-replayable} sequence.  We use the $T$ entries preceding the test item as a warmup window ($T=1$ in main experiments), running the full retrieve-and-rerank pipeline on this window under the current $\phi_u$ and repurposing $H_u$ as a \emph{per-user self-supervised signal}:
\begin{equation}
\mathcal{R}_u \;=\; \bigl\{(C_j,\,\hat y_j,\,y_j)\bigr\}_{j=1}^{T}, \qquad \mathcal{R}_u^{\mathrm{hit}}:=\{j : \hat y_j=y_j\},\;\mathcal{R}_u^{\mathrm{miss}}:=\{j : \hat y_j\ne y_j\}.
\label{eq:record}
\end{equation}
The hit set characterises effective profile coverage; the miss set pinpoints systematic blind spots.  Conditioned on $\mathcal{R}_u$, a single LLM reflection call uses a structured Chain-of-Thought template~\cite{wei2022chain} to propose a candidate profile update:
\begin{equation}
\tilde\phi_u \;=\; \mathrm{Reflect}\bigl(\mathcal{R}_u^{\mathrm{hit}},\,\mathcal{R}_u^{\mathrm{miss}},\,\phi_u\bigr).
\label{eq:evolve}
\end{equation}
This call fires only once per user, builds $\phi_u$ when the user has no profile yet, and does not touch other pipeline components.

\paragraph{Critic gate.}
Because chain-of-thought reasoning can be unfaithful~\cite{sun2026fact}, an \emph{acceptance gate} can optionally be imposed atop the reflection: the critic receives the old profile $\phi_u$, the proposed $\tilde\phi_u$, and the warmup record $\mathcal{R}_u$, returning a binary decision $\hat s(\tilde\phi_u;\phi_u)\in\{\mathrm{accept},\mathrm{reject}\}$:
\begin{equation}
\phi_u^{(1)} \;=\;
\begin{cases}
\tilde\phi_u, & \text{if}\ \hat s(\tilde\phi_u;\,\phi_u)=\mathrm{accept},\\
\phi_u,       & \text{otherwise}.
\end{cases}
\label{eq:accept_rule}
\end{equation}
The critic is a conservative safeguard: it keeps the old profile when the proposal is clearly less supported by the warmup cases, which prevents the reflective rewrite from pulling the user profile away from $H_u$.  We use a \emph{zero-extra-LLM} critic, a deterministic check of how well the proposed profile covers the warmup target items, so the gate's marginal cost over $\mathrm{Reflect}$ is negligible (Appendix~\ref{app:prompts}).  Each user sees only one pass of Eq.~\ref{eq:accept_rule} (C3): additional passes would multiply the LLM cost while reusing the same warmup record $\mathcal{R}_u$.

\subsection{Operator Separability and Ablation Predictions}
\label{sec:failure_modes_ref}

The three operators of \S\ref{sec:profile}--\S\ref{sec:evolution} are designed to be \emph{separable}: each addresses one structural challenge that its ablation re-introduces.  \textbf{Vocabulary mismatch} (addressed by \S\ref{sec:profile}): when the item has no profile ($\phi_i\equiv\varnothing$), there is no second vocabulary to align against; \textbf{Channel capture} (addressed by \S\ref{sec:communication}): prompt-injected $\phi_u$ competes with $M_{\mathrm{collab}}$ for the reranker budget $B$ (C1), so any profile-content gain is capped by the collaborative-memory space yielded; \textbf{Communicative drift} (addressed by \S\ref{sec:evolution}): profiles written at warmup are not calibrated against warmup feedback, and the update of Eq.~\ref{eq:accept_rule} performs this calibration once.  Full pseudocode, per-case traces, and complexity analysis are in Appendix~\ref{app:algorithm}, \ref{app:cases} and~\ref{app:cost}; each row of the main ablation (\S\ref{sec:experiments}, Table~\ref{tab:ablation}) is labelled by the challenge it isolates, so the whole table can be read as a test of the three operators' separability.

\section{Experiments}
\label{sec:experiments}

\subsection{Experimental Setup}

\paragraph{Datasets and baselines.} We evaluate on four InstructRec~\cite{xu2025iagent} benchmarks: \textbf{Amazon Books} (7.4K users, 120.9K items; abbreviated \emph{Books}), \textbf{Goodreads} (11.7K / 57.4K), \textbf{MovieTV} (5.6K / 29.0K), \textbf{Yelp} (3.0K / 31.6K), using the original splits and natural-language user instructions.  Baselines span three families: traditional recommenders (LightGCN~\cite{he2020lightgcn}, SASRec~\cite{kang2018self}, P5~\cite{geng2022recommendation}); single-sided LLM/agentic methods (Vanilla LLM~\cite{liu2023chatgpt}, iAgent~\cite{xu2025iagent}, AgentCF~\cite{zhang2024agentcf}, RecBot~\cite{tang2026interactive}, i$^2$Agent~\cite{xu2025iagent}); and memory-augmented agentic methods, whose strongest published representative on these splits is MemRec~\cite{chen2026memrec}.  All baseline numbers are imported from the originating papers ($^\dagger$ in Table~\ref{tab:main_all}); per-dataset characteristics and baseline details are in Appendix~\ref{app:dataset_details} and~\ref{app:baseline_details}.

\paragraph{\ours{} configuration.}
\ours{} is instantiated on a frozen \textbf{gpt-4o-mini} backbone and runs a single round of profile warmup per user.  The \textsc{Understand} operator uses homophily floors $\alpha_{ui}{=}0.4$ on user--item edges and $\alpha_{uu}{=}0.3$ on user--user edges, with a preference-dominant channel weighting $(w_c)$.  The \textsc{Sync} operator runs $T_{\mathrm{evo}}{=}1$ reflective pass per user (conditioned on the second-to-last interaction) optionally filtered by the critic.  The retrieval budget $k$ is chosen on a development split from $\{4, 8, 16\}$ and fixed at $k{=}16$ for all main-table numbers.  Hit@1 improvements are tested by per-dataset one-sided binomial tests with Wilson 95\% CIs (Appendix~\ref{app:significance}); the cross-dataset joint null is controlled by Holm--Bonferroni at family-wise $\alpha{=}0.05$; ladder transitions use exact paired McNemar. Our code is available at https://github.com/Jove-Zhen/HARP.

\subsection{Main Results}

\begin{table*}[t]
\centering
\caption{Main results on four benchmarks.  Bold $=$ best, underline $=$ second best; $^\dagger$ = from original paper.  ``Rel.\ Improv.''\ = relative improvement over the strongest published baseline.  All H@1 improvements $p<10^{-9}$ (Holm--Bonferroni corrected).}
\label{tab:main_all}
\resizebox{\textwidth}{!}{%
\begin{tabular}{ll|rrrrr|rrrrr}
\toprule
& & \multicolumn{5}{c|}{\textbf{Books}} & \multicolumn{5}{c}{\textbf{Goodreads}} \\
\textbf{Paradigm} & \textbf{Model} & H@1 & H@3 & N@3 & H@5 & N@5 & H@1 & H@3 & N@3 & H@5 & N@5 \\
\midrule
\multirow{3}{*}{\rotatebox{90}{\scriptsize Trad.}}
& LightGCN$^\dagger$ & .175 & .326 & .260 & .570 & .359 & .250 & .588 & .443 & .790 & .526 \\
& SASRec$^\dagger$ & .091 & .283 & .200 & .485 & .282 & .132 & .352 & .258 & .541 & .335 \\
& P5$^\dagger$ & .219 & .361 & .299 & .527 & .367 & .157 & .323 & .251 & .506 & .326 \\
\midrule
\multirow{5}{*}{\rotatebox{90}{\scriptsize Agent (match)}}
& Vanilla LLM$^\dagger$ & .314 & .562 & .453 & .727 & .523 & .286 & .466 & .395 & .739 & .504 \\
& iAgent$^\dagger$ & .393 & .556 & .486 & .691 & .541 & .262 & .495 & .395 & .659 & .463 \\
& RecBot$^\dagger$ & .398 & .549 & .485 & .679 & .538 & .271 & .475 & .388 & .650 & .459 \\
& AgentCF$^\dagger$ & .346 & .606 & .496 & .740 & .551 & .295 & .591 & .465 & .773 & .540 \\
& i$^2$Agent$^\dagger$ & .445 & .652 & .565 & .771 & .614 & .310 & .608 & .483 & .768 & .548 \\
\midrule
{\scriptsize Mem.-aug.}
& MemRec$^\dagger$ & \underline{.512} & \underline{.692} & \underline{.615} & \underline{.801} & \underline{.660} & \underline{.400} & \underline{.666} & \underline{.554} & \underline{.805} & \underline{.611} \\
\midrule
\rowcolor{blue!6}
{\scriptsize \textbf{Comm.}}
& \textbf{\ours{}} & \textbf{.547} & \textbf{.720} & \textbf{.647} & \textbf{.819} & \textbf{.688} & \textbf{.469} & \textbf{.715} & \textbf{.612} & \textbf{.847} & \textbf{.664} \\
\rowcolor{blue!6}
& \textit{Rel.\ Improv.} & \textit{+6.9\%} & \textit{+4.2\%} & \textit{+5.2\%} & \textit{+2.2\%} & \textit{+4.2\%} & \textit{+17.2\%} & \textit{+7.4\%} & \textit{+10.4\%} & \textit{+5.2\%} & \textit{+8.6\%} \\
\bottomrule
\end{tabular}%
}

\vspace{4pt}

\resizebox{\textwidth}{!}{%
\begin{tabular}{ll|rrrrr|rrrrr}
\toprule
& & \multicolumn{5}{c|}{\textbf{MovieTV}} & \multicolumn{5}{c}{\textbf{Yelp}} \\
\textbf{Paradigm} & \textbf{Model} & H@1 & H@3 & N@3 & H@5 & N@5 & H@1 & H@3 & N@3 & H@5 & N@5 \\
\midrule
\multirow{3}{*}{\rotatebox{90}{\scriptsize Trad.}}
& LightGCN$^\dagger$ & .348 & .564 & .474 & .688 & .524 & .344 & .566 & .472 & .755 & .549 \\
& SASRec$^\dagger$ & .340 & .523 & .447 & .638 & .494 & .231 & .431 & .346 & .560 & .398 \\
& P5$^\dagger$ & .170 & .321 & .255 & .501 & .329 & .144 & .321 & .244 & .522 & .479 \\
\midrule
\multirow{5}{*}{\rotatebox{90}{\scriptsize Agent (match)}}
& Vanilla LLM$^\dagger$ & .405 & .776 & .610 & .860 & .645 & .169 & .528 & .370 & .686 & .436 \\
& iAgent$^\dagger$ & .425 & .617 & .536 & .742 & .587 & .400 & .601 & .515 & .730 & .568 \\
& RecBot$^\dagger$ & .437 & .611 & .538 & .731 & .587 & .401 & .600 & .516 & .717 & .564 \\
& AgentCF$^\dagger$ & .391 & .669 & .552 & .786 & .601 & .193 & .437 & .333 & .637 & .415 \\
& i$^2$Agent$^\dagger$ & .491 & .723 & .626 & .822 & .667 & .421 & .645 & .552 & .765 & .601 \\
\midrule
{\scriptsize Mem.-aug.}
& MemRec$^\dagger$ & \underline{.588} & \underline{.782} & \underline{.701} & \underline{.882} & \underline{.742} & \underline{.487} & \underline{.691} & \underline{.605} & \underline{.791} & \underline{.646} \\
\midrule
\rowcolor{blue!6}
{\scriptsize \textbf{Comm.}}
& \textbf{\ours{}} & \textbf{.645} & \textbf{.838} & \textbf{.758} & \textbf{.917} & \textbf{.790} & \textbf{.604} & \textbf{.790} & \textbf{.712} & \textbf{.865} & \textbf{.743} \\
\rowcolor{blue!6}
& \textit{Rel.\ Improv.} & \textit{+9.7\%} & \textit{+7.2\%} & \textit{+8.1\%} & \textit{+4.0\%} & \textit{+6.5\%} & \textit{+24.0\%} & \textit{+14.3\%} & \textit{+17.7\%} & \textit{+9.4\%} & \textit{+15.0\%} \\
\bottomrule
\end{tabular}%
}
\end{table*}

Table~\ref{tab:main_all} reports the complete results.  \ours{} consistently outperforms all baselines on every dataset and every metric.  Three patterns are worth noting.  \emph{First}, the dominance holds across all three baseline families: Hit@1 over the strongest traditional baseline is $1.8\times$--$2.5\times$, over the strongest single-sided agent i$^2$Agent it is $+23\%$/$+51\%$/$+31\%$/$+43\%$ on Books/Goodreads/MovieTV/Yelp, and over the strongest memory-augmented baseline MemRec it ranges from $+6.9\%$ (Books) to $+24.0\%$ (Yelp).  \emph{Second}, the gain is largest on Yelp and Goodreads ($+24.0\%$ / $+17.2\%$), consistent with \ours{}'s profile-based re-weighting of neighbours exactly where the collaborative-only signal is weakest.  \emph{Third}, the relative gain over MemRec shrinks monotonically with $k$ (e.g.\ Yelp $+24.0\% \to +14.3\% \to +9.4\%$ at Hit@$\{1,3,5\}$), indicating that \ours{} promotes the ground truth \emph{into} the top-1 slot rather than merely reshuffling an already-correct pool.  On the full test splits, the Wilson 95\% CI lower bound of \ours{}'s Hit@1 strictly dominates the MemRec point estimate with margins $+2.38/+5.98/+4.38/+9.93$ pp on Books/Goodreads/MovieTV/Yelp; per-dataset one-sided binomial $p<10^{-9}$ and Holm--Bonferroni adjusted $p<10^{-9}$ (Appendix~\ref{app:significance}).

\subsection{Ablation: Which Components of \ours{} Carry the Gain?}

\begin{figure}[t]
\centering
\includegraphics[width=0.95\linewidth]{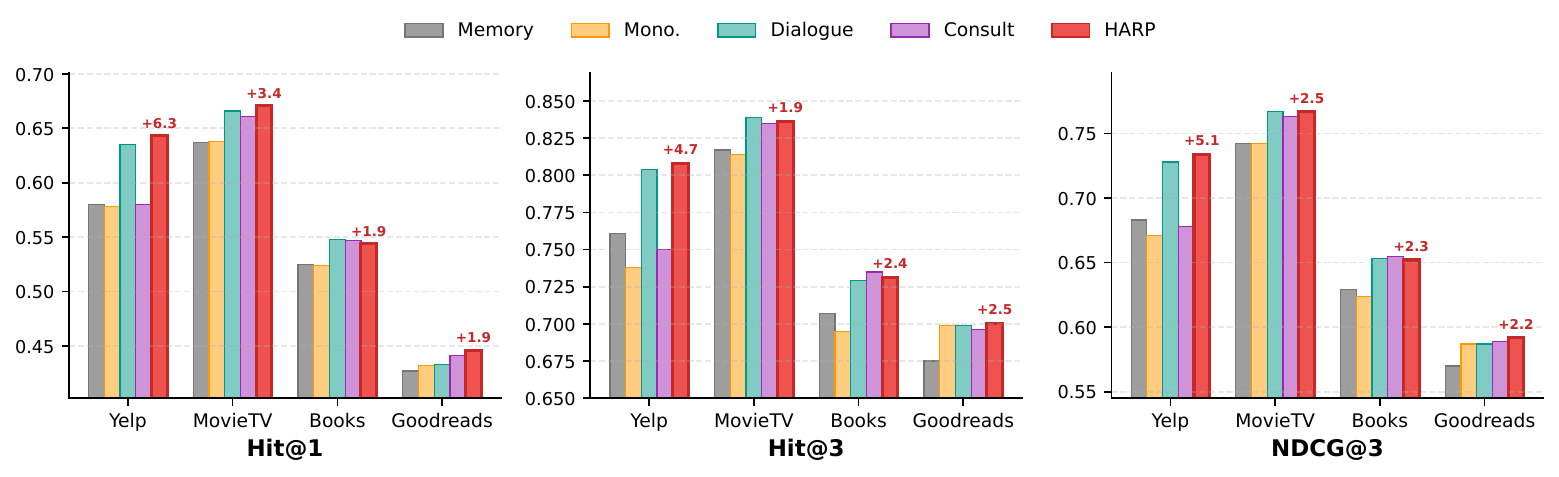}
\caption{Five-rung ablation ladder (four benchmarks, matched $1{,}000$-user subsplits).  (a)~Hit@1; (b)~Hit@3; (c)~NDCG@3.  Red numbers = total gain (pp) over \textsc{Memory}.}
\label{tab:ladder_cross}
\end{figure}

To systematically attribute contributions, we define five ablation configurations sharing the same frozen backbone and differing only in ``which \ours{} components are enabled'': \textsc{Memory} ($\phi_u{=}\phi_i{=}\varnothing$; no gating, no reflection); \textsc{Monologue} (user profile $\phi_u$ injected into the reranker prompt only); \textsc{Dialogue} (both $\phi_u,\phi_i$ enabled, homophily gating $\mathcal{H}$ used on retrieval; \emph{no} prompt injection, \emph{no} reflection); \textsc{Consult} (both profiles but \emph{retrieval-side gate disabled}; a direct ``retrieval-side vs.\ prompt-side'' contrast against \textsc{Dialogue}); \textbf{\ours{}} (full pipeline: both profiles, retrieval-side gate, prompt injection, one-pass reflection).  Table~\ref{tab:ablation} reports the ablation on the $1{,}000$-user Yelp subsplit; Figure~\ref{tab:ladder_cross} replicates the same five rungs across all four benchmarks.

\begin{table}[t]
\centering
\small
\caption{\ours{} ablation (Yelp subsplit).  CH = retrieval-side homophily gate; RI = reranker prompt injection; Sync = reflection.}
\label{tab:ablation}
\begin{tabular}{l|ccccc|ccccc}
\toprule
\textbf{Config} & $\phi_u$ & $\phi_i$ & \textbf{CH} & \textbf{RI} & \textbf{Sync} & \textbf{H@1} & \textbf{H@3} & \textbf{H@5} & \textbf{N@3} & \textbf{N@5} \\
\midrule
\textsc{Memory}    &   &   &   &   &   & .580 & .761 & .855 & .683 & .722 \\
\textsc{Monologue} & \checkmark &   &   & \checkmark &   & .578 & .738 & .824 & .671 & .707 \\
\textsc{Consult}   & \checkmark & \checkmark &   & \checkmark &   & .580 & .750 & .842 & .678 & .716 \\
\textsc{Dialogue}  & \checkmark & \checkmark & \checkmark &   &   & .635 & .804 & .888 & .728 & .763 \\
\rowcolor{blue!5}
\textbf{\ours{}}   & \checkmark & \checkmark & \checkmark & \checkmark & \checkmark & \textbf{.643} & \textbf{.808} & \textbf{.889} & \textbf{.734} & \textbf{.767} \\
\bottomrule
\end{tabular}
\end{table}

\paragraph{Three key findings.}
\textbf{(1) The retrieval-side gate is the primary source of gain:} \textsc{Monologue}$\to$\textsc{Dialogue} jumps H@1 by $+5.7$ pp; prompt-side injection alone (\textsc{Consult} vs.\ \textsc{Memory}) does not move H@1.  \textbf{(2) The gain needs the profiles to enter the gate:} \textsc{Monologue} (user profile only) and \textsc{Consult} (both profiles, no gate) stay at $.578$--$.580$, the same as \textsc{Memory}; only \textsc{Dialogue}, which uses both profiles in the retrieval-side gate, improves.  \textbf{(3) Prompt injection and \textsc{Sync} add a small residual:} \textsc{Dialogue}$\to$\ours{} adds $+0.8$ pp H@1 for the two together (prompt injection alone does not move H@1).  Paired McNemar (Appendix~\ref{app:pairwise}) confirms \textsc{Monologue}$\to$\textsc{Dialogue} ($p{=}7.2\!\times\!10^{-6}$) and \textsc{Consult}$\to$\ours{} ($p{=}2.5\!\times\!10^{-7}$) are significant, \textsc{Memory}$\to$\textsc{Monologue} lies within sampling noise ($p{=}0.94$), and \textsc{Dialogue}$\to$\textsc{Consult} is significantly negative ($-5.5$ pp, $p{=}1.0\!\times\!10^{-5}$), which supports consuming the profile on the retrieval side.

\paragraph{Do the ablation conclusions hold across datasets?}
Figure~\ref{tab:ladder_cross} shows the five-rung ladder on all four benchmarks (matched $1{,}000$-user subsplits) for three representative metrics: Hit@1, Hit@3, and NDCG@3.  The conclusion is robust: \textsc{Monologue}\,$\approx$\,\textsc{Memory}, \textsc{Consult}\,$\approx$\,\textsc{Memory}, \textsc{Dialogue}\,$>$\,\textsc{Monologue} is the largest single-rung jump on Yelp, MovieTV and Books, and the \textsc{Dialogue}$\to$\ours{} residual (injection and \textsc{Sync} together) is positive on Yelp/MovieTV/Goodreads ($+0.8/+0.5/+1.3$ pp Hit@1) and slightly negative on Books ($-0.4$ pp).  Net Hit@1 gains over \textsc{Memory}: $+6.3/+3.4/+1.9/+1.9$ pp (Yelp/MovieTV/Books/Goodreads); a per-user repair-source decomposition (Appendix~\ref{app:ha_bucket_empirical}) confirms that the majority of L0-miss$\to$L4-hit transitions are already present at the \textsc{Dialogue} stage, attributing the primary engine of improvement to the retrieval-side gate.  The four-metric ladder line plots are in Appendix Figure~\ref{fig:ladder_lines}.

\paragraph{Repair sources and the Sync trade-off.}
To further understand the mechanism behind the gains, we perform a per-user outcome-transition decomposition on three deep-analysis subsplits (Yelp/MovieTV/Books, each $1{,}000$ users): each user is tagged by the triple $(L_0,L_2,L_4)\in\{0,1\}^3$ indicating hit/miss under \textsc{Memory}, \textsc{Dialogue}, and \ours{}.  The net repair counts over \textsc{Memory} are $+63/+34/+19$ users ($+6.3/+3.4/+1.9$ pp), with the ratio of retrieval-gate repairs (Dial-rep) to \textsc{Dialogue}$\to$\ours{} repairs (Sync-rep, injection and \textsc{Sync} together) stable at $5.3{:}1$ (Yelp), $3.2{:}1$ (MovieTV), $3.0{:}1$ (Books).  This confirms that the \textsc{Dialogue}-stage homophily gate is the \emph{primary engine} of improvement, with the \textsc{Dialogue}$\to$\ours{} step (injection and \textsc{Sync}) providing marginal gains on the sub-population that \textsc{Dialogue} alone cannot reach. Notably, this step also introduces a small number of regression events ($-0.4$ pp on Books), reflecting a trade-off between conservatism and coverage; three qualitative case studies in Appendix~\ref{app:cases} examine it.  Furthermore, the Pearson correlation between a profile-length proxy and per-user gain is $\rho\in[-0.048,-0.005]$, confirming that reflection reshapes the profile's \emph{composition} rather than its length.

\section{Conclusion}
\label{sec:conclusion}

We recast LLM-based recommendation from an \emph{agentic-communication} perspective in which user and item agents exchange structured self-descriptions along a four-stage process (\textsc{Build}--\textsc{Transmit}--\textsc{Understand}--\textsc{Sync}), use the formulation analytically on the dominant memory-augmented agentic paradigm to identify three open communication challenges (asymmetric \textsc{Build}, bandwidth-trapped \textsc{Understand}, and absent \textsc{Sync}), and introduce \ours{}, a three-operator training-free instantiation on a frozen LLM backbone: a symmetric item profile (\textsc{Build}), a channel-decomposable homophily-gated retrieval (\textsc{Understand}), and a single-pass reflective update (\textsc{Sync}).  On four public benchmarks, \ours{} lifts Hit@1 by $+6.9$--$24.0\%$ over the strongest published baseline, and ablations further confirm that operator-level gains align with the three challenges.

\bibliography{references}
\bibliographystyle{plainnat}

\appendix

\section{Extended Related Work}
\label{app:rw_extended}

\paragraph{Multi-agent communication and skill-based self-improvement.}
A rich literature studies how artificial agents develop communication protocols~\cite{foerster2016learning, lazaridou2016multi, mordatch2018emergence, ren2020compositional}, and recent work extends this to LLM-based agents communicating in natural language~\cite{li2023camel, hong2024metagpt, du2023improving}, typically in multi-round dialogues.  A parallel line shows that LLM agents can accumulate reusable skills through experience~\cite{wang2023voyager, zhu2023ghost, madaan2023self, shinn2023reflexion}: Voyager builds a skill library in Minecraft, GITM decomposes tasks into learnable skills, and self-refine approaches demonstrate iterative LLM self-improvement.  \ours{} uses both lenses orthogonally: user and item agents exchange one-shot structured messages measured by ranking accuracy rather than dialogue fluency, and the reflective update operates on the agent's \emph{profile} under a acceptance rule (Eq.~\ref{eq:accept_rule}) rather than accumulating a skill library.  The reranker-prompt bandwidth constraint (C1 in \S\ref{sec:overview}) and the channel decomposability of Remark~\ref{thm:decomp} matter most in this one-shot, high-throughput setting.

\paragraph{LLM-based agents for recommendation (detailed thread).}
A growing body treats recommendation as an LLM-agent task~\cite{zhang2024agentcf, huang2025recommender, wang2024recmind, friedman2023leveraging, xu2025iagent, tang2026interactive}: AgentCF~\cite{zhang2024agentcf} pioneers the user/item-as-agent formulation with free-text memories on both sides, but neither side carries a structured, channel-aligned profile; InteRecAgent~\cite{huang2025recommender} introduces tool-augmented user agents; iAgent~\cite{xu2025iagent} keeps a static per-user profile with no reflective update; i$^2$Agent~\cite{xu2025iagent} adds a per-user reflection loop but \emph{without} an acceptance rule, so every proposed revision is applied; RecBot~\cite{tang2026interactive} supports session-level dynamic memory; RecAgent~\cite{wang2025user} simulates user behaviour.  Relative to this line, three design choices in \ours{} are jointly novel: (i)~a structured profile on the \emph{item} side with the same four-channel schema as the user side (no prior work gives the item agent a structured, channel-aligned profile), (ii)~a homophily signal consumed during \emph{retrieval} rather than inside the reranker prompt, and (iii)~an acceptance check on the reflective update (self-reflective predecessors apply every proposed revision).  A parallel line develops general agentic memory architectures beyond recommendation (A-MEM~\cite{xu2026mem}, Memory-R1~\cite{yan2026memory}, Zep~\cite{rasmussen2025zep}, H-Mem~\cite{ye2026hmem}) offering infrastructure pluggable at the \textsc{Transmit} stage; a recent dynamic-graph thread~\cite{tao2026task} makes retrieval augmentation itself task-aware, conceptually close to our \textsc{Understand} routing.

\paragraph{Tuning- and grounding-based LLM recommendation.}
A parallel line tunes or grounds the LLM against the recommendation objective rather than operating it as an agent: TALLRec~\cite{bao2023tallrec}, CoLLM~\cite{zhang2025collm}, LLaRA~\cite{liao2024llara}, BIGRec~\cite{bao2025bi}, A-LLMRec~\cite{kim2024large}, LLMRec~\cite{wei2024llmrec}, and ONCE~\cite{liu2024once}; surveys~\cite{wu2024survey} place these on a spectrum of \emph{what} in the LLM is adapted.  \ours{} is complementary and sits at the \emph{interaction-loop} end: it keeps the LLM frozen, leaves the collaborative signal in place, and restructures \emph{how} user and item agents exchange profiles along the pipeline; it can in principle be layered on top of any of these backbones.  The retrieval-vs-prompt rationale of \S\ref{sec:communication} is motivated by adaptive retrieval in retrieval-augmented generation~\cite{asai2024self,jiang2023active}, where evidence is selected before generation; its support in our setting is the ablation of \S\ref{sec:experiments}.

\paragraph{Relation to adjacent LLM-agent currents.}
Three concurrent LLM-agent currents intersect with \ours{} without overlapping its main contribution: the \emph{self-evolving-agent} line~\cite{gao2025survey,zhang2026agentic} (replacing fine-tuning with in-context rewrites; our \textsc{Sync} is one such primitive but we do not claim a new mechanism), the \emph{agentic-context-engineering} view~\cite{zhang2026agentic} (our dual-profile schema, Table~\ref{tab:channels}, is a user/item-symmetric realisation specific to recommendation), and \emph{agentic RAG}~\cite{singh2025agentic} (our homophily-gated score, Eq.~\ref{eq:comm_score}, is a retrieval decision gated by a profile-level homophily rather than a learned policy).  Channel-wise decomposition of graph homophily itself is not new: multi-relational GNNs~\cite{schlichtkrull2018modeling,hu2020heterogeneous} and per-class edge-homophily~\cite{zhu2020beyond} already exploit the idea at the interaction / class-label level.  Relative to those threads, \ours{} lifts the decomposition to the \emph{profile} level (agent-authored, agent-mutable), and the three-challenge analysis of the four-stage Build--Transmit--Understand--Sync formulation is the contribution specific to agentic recommendation.

\section{Experimental Setup Details}
\label{app:setup_details}

\subsection{Dataset Details}
\label{app:dataset_details}

We evaluate on four public datasets widely used in recommendation research, used under their original terms of use for research, encompassing diverse domains such as e-commerce, social reading, entertainment, and local services. As stated in \S\ref{sec:experiments}, we adopt the versions of these datasets augmented with natural-language user instructions from \textbf{InstructRec}~\cite{xu2025iagent}. The original data sources and their characteristics are summarised below.

\paragraph{Books.} Derived from the \textbf{Amazon review dataset}~\cite{ni2019justifying}, this subset focuses on book recommendations. It is characterised by extremely sparse interactions and a very large item space; user preferences in this domain are typically stable and highly content-driven, centred on specific genres, authors, or themes.

\paragraph{Goodreads.} Collected from the \textbf{Goodreads} social book-cataloguing website~\cite{wan2019fine}, this dataset is comparatively dense. It features strong community interactions and rich metadata about books, including series information, and users often exhibit series-aware reading behaviours and are influenced by social signals.

\paragraph{MovieTV.} Also originating from the \textbf{Amazon review dataset}~\cite{ni2019justifying}, this dataset covers movies and TV shows. The domain is marked by volatile user preferences often influenced by immediate context or trending content; while metadata such as genre and cast are important, item recency frequently plays a critical role in user decision-making.

\paragraph{Yelp.} Sourced from the \textbf{Yelp Dataset}, this dataset consists of reviews for local businesses such as restaurants and services. It is characterised by strong categorical constraints (e.g., cuisine type) and by the critical importance of attributes like price range and location; user preferences are often highly context-dependent.

\subsection{Baseline Details}
\label{app:baseline_details}

This appendix provides detailed descriptions of the baseline models used in our comparative evaluation in \S\ref{sec:experiments}. Following the categorisation in the main text, we group the baselines into three families.

\paragraph{(I) Traditional recommenders.} These models mostly represent the pre-LLM paradigm (P5 being the exception), in which historical information is encoded as dense latent vectors without explicit semantic memory.
\begin{itemize}[leftmargin=*, itemsep=2pt, topsep=2pt]
    \item \textbf{LightGCN}~\cite{he2020lightgcn}: a graph-collaborative-filtering model that simplifies the GCN design by removing feature transformation and nonlinear activation, learning user/item embeddings via linear propagation on the user--item interaction graph.
    \item \textbf{SASRec}~\cite{kang2018self}: a self-attentive sequential recommender that models the entire user sequence to capture long-term and dynamic dependencies over the history.
    \item \textbf{P5}~\cite{geng2022recommendation}: a unified framework that formulates recommendation tasks as sequence-to-sequence language modelling over a pre-trained T5 backbone, representing users and items as sequence tokens within personalised prompts.
\end{itemize}

\paragraph{(II) Single-sided LLM / agentic methods.} These approaches operate on raw interaction histories or static profiles without maintaining a persistent, structured, and evolving semantic memory.
\begin{itemize}[leftmargin=*, itemsep=2pt, topsep=2pt]
    \item \textbf{Vanilla LLM}~\cite{liu2023chatgpt}: direct zero-shot prompting of an instruction-tuned LLM (\texttt{gpt-4o-mini} in our protocol), in which the user's history is serialised into a natural-language context and the LLM selects from the candidate list; a baseline measuring the LLM's inherent capability without any memory design.
    \item \textbf{iAgent}~\cite{xu2025iagent}: an LLM agent with a static per-user profile constructed from historical interactions and descriptive data; the profile is fed into the LLM as context but does not adapt over time after the initial construction.
    \item \textbf{i$^2$Agent}~\cite{xu2025iagent}: an extension of \textbf{iAgent} that adds a self-reflection mechanism, allowing the agent to refine its internal state after recommendations, with reflections confined to the individual agent's experience with a specific user.
    \item \textbf{AgentCF}~\cite{zhang2024agentcf}: an agent-based collaborative-filtering framework in which user- and item-agents interact autonomously and update their own preferences/characteristics, with memory updates localised to the interacting agents.
    \item \textbf{RecBot}~\cite{tang2026interactive}: a conversational recommender in which an LLM maintains a dynamic dialogue history and updates user-preference estimates during multi-turn sessions; dynamics are bounded to the current session.
\end{itemize}

\paragraph{(III) Memory-augmented agentic methods.}
\begin{itemize}[leftmargin=*, itemsep=2pt, topsep=2pt]
    \item \textbf{MemRec}~\cite{chen2026memrec}: the strongest published memory-augmented agentic recommender on these InstructRec splits, and our primary end-to-end baseline.
\end{itemize}

\subsection{Reproducibility Configuration}
\label{app:repro_checklist}

Section~\ref{sec:experiments} and Appendix~\ref{app:hyper} specify the LLM endpoint, dataset splits, rule-extractor artefacts, hyperparameters and significance protocol.  Code is available at \url{https://github.com/Jove-Zhen/HARP}.  This subsection covers two additional operational concerns.

\paragraph{Critic configuration.}
The critic of Eq.~\ref{eq:accept_rule} produces a binary \{accept, reject\} decision per user.  It is a conservative, optional component (\S\ref{sec:evolution}).

\paragraph{Per-user randomness control.}
We use a single seed (\texttt{seed=42}) throughout the paper for user sampling and candidate construction.  LLM decoding and parallel execution may still cause small run-to-run variation.

We formally collect the statements and proofs invoked throughout the
main text.  Appendix~\ref{app:proof_decomp} proves
Remark~\ref{thm:decomp} (channel decomposability of the profile-level
homophily) in full, with all intermediate steps made explicit.
Appendix~\ref{app:lemma_ablation} states and proves
Lemma~\ref{lem:channel_ablation} (channel-ablation identity), a direct
consequence that underwrites the per-channel contribution decomposition
reported in Appendix~\ref{app:ha_bucket_empirical}.
Appendix~\ref{app:prop_budget} records the retrieval-vs-prompt
design rationale of \S\ref{sec:communication} as a non-formal remark;
the rationale is tested directly by the table-level measurements of
\S\ref{sec:experiments} (Table~\ref{tab:ablation}) rather than lifted
to a theorem.

\section{Proofs}
\label{app:proofs_top}

\subsection{Proof of Remark~\ref{thm:decomp} (Channel Decomposability)}
\label{app:proof_decomp}

\paragraph{Setup.}
Let $\mathcal{C}$ be the finite set of semantic channels
(Table~\ref{tab:channels}) and represent each profile $\phi$ as a typed
tuple $(\phi^{(c)})_{c \in \mathcal{C}}$ of per-channel phrase sets,
with non-negative channel weights $(w_c)_{c\in\mathcal{C}}$,
$\sum_c w_c = 1$ (Appendix~\ref{app:hyper}).  Let
$\mathcal{H}_c(\phi_u,\phi_i)$ denote the per-channel soft overlap
(\texttt{preference}/\texttt{audience} channels: phrase-coverage ratio;
\texttt{quality}/\texttt{context} channels: bounded rule-based
indicators in $[0,1]$; see Appendix~\ref{app:item_extractor}), and
write $\mathcal{H}(\phi_u,\phi_i) = \sum_{c\in\mathcal{C}} w_c\,
\mathcal{H}_c(\phi_u,\phi_i)$ as in Eq.~\ref{eq:homophily_def}.

\paragraph{Claim and proof.}
By step R4 of \S\ref{sec:profile}, every surface phrase is routed to
\emph{at most one} channel, so each $\mathcal{H}_c$ is a function of
the channel-$c$ projections alone.  Combined with the linear weighted
sum, $\partial\mathcal{H}/\partial\mathcal{H}_c = w_c$ for every $c$,
and the Hessian of the map
$(\mathcal{H}_c)_{c\in\mathcal{C}}\mapsto\mathcal{H}$ vanishes; hence
perturbing any subset $S\subseteq\mathcal{C}$ of channels to
$\mathcal{H}_c = 0$ shifts $\mathcal{H}$ by exactly
$-\sum_{c\in S} w_c\,\mathcal{H}_c$, with no cross-channel coupling.
$\square$

\paragraph{Scope.}
This is a decomposability property \emph{of $\mathcal{H}$ itself}, not
of the reranker's downstream decision.  It underwrites per-channel
inspection of the homophily quantity; how channels combine through the
reranker to affect Hit@1 is an empirical object studied in
\S\ref{sec:experiments}.  The boundary case $|\mathcal{C}|=1$ with
$\mathrm{sem}$ reduced to $\mathbf{1}\{\cdot{=}\cdot\}$ reduces to the
per-edge same-class indicator, whose average over edges is the scalar edge-homophily ratio
$h_{\mathrm{edge}}=|E^{=}|/|E|$ of~\cite{zhu2020beyond}.

\subsection{Lemma~\ref{lem:channel_ablation} (Channel-Ablation Identity)}
\label{app:lemma_ablation}

\begin{lemma}[Channel-ablation identity]
\label{lem:channel_ablation}
For any subset $S\subseteq\mathcal{C}$ of channels, let
$\mathcal{H}^{(-S)}(\phi_u,\phi_i)$ denote the homophily computed with
$\mathcal{H}_c\gets 0$ for every $c\in S$.  Then
\begin{equation}
\mathcal{H}(\phi_u,\phi_i) \;-\; \mathcal{H}^{(-S)}(\phi_u,\phi_i)
\;=\; \sum_{c\in S} w_c\,\mathcal{H}_c(\phi_u,\phi_i),
\label{eq:ablation_identity}
\end{equation}
and the drop is strictly positive whenever there exists $c\in S$ with
$w_c>0$ and $\mathcal{H}_c>0$.
\end{lemma}

\paragraph{Proof.}
Direct consequence of Appendix~\ref{app:proof_decomp}; strict
positivity follows because each summand in the sum is non-negative.
$\square$

\paragraph{Scope.}
The identity is a statement about $\mathcal{H}$, not about Hit@1:
the per-channel quantities $w_c\,\mathcal{H}_c$ sum exactly back to
$\mathcal{H}$, but how they translate into changes in reranker-level
Hit@1 depends on the reranker's response function and is measured
empirically in \S\ref{sec:experiments}.

\subsection{Retrieval--Prompt Design Rationale}
\label{app:prop_budget}

\begin{remark}[Retrieval--prompt design rationale]
\label{prop:budget}
Under a fixed reranker context budget $B$ and a frozen LLM backbone
(constraints (C1)--(C2) of \S\ref{sec:overview}), injecting the user
profile $\phi_u$ on the retrieval side (Eq.~\ref{eq:comm_score}) is
preferable to injecting it on the reranker-prompt side
(Eq.~\ref{eq:rerank}), for two reasons:
\begin{itemize}[leftmargin=*,itemsep=1pt,topsep=2pt]
    \item \emph{Budget competition.} Once the budget is saturated, every prompt-side profile token trades against a token of the collaborative-memory block
    $M_{\mathrm{collab}}$, which is the most information-dense block in
    the prompt; prompt-side injection is therefore a zero-sum trade
    rather than a free addition.
    \item \emph{Irreversibility.} The reranker reads only the
    memories of the candidates admitted at retrieval time; once
    misleading candidates fill $M_{\mathrm{collab}}$, prompt-side
    profile information cannot replace them.
\end{itemize}
We treat this as a \emph{design rationale} rather than a theorem: its
direction-of-change is consistent with our data
(Table~\ref{tab:ablation}: retrieval-side $+5.5$\,pp vs.\ prompt-side
$\le 0.2$\,pp on $1{,}000$-user Yelp), but its magnitude is a property
of the specific dataset/backbone and is not claimed to extrapolate.
\end{remark}

\vspace{10pt}

We turn from derivations to the empirical measurements that accompany
the methodology of \S\ref{sec:methodology}.

\paragraph{Reading map for this part.}
The subsections below do not report novel findings in isolation;
each instead answers one class of potential objection to the main
\ours{} results.  Concretely:

\begin{itemize}[leftmargin=*,itemsep=2pt,topsep=2pt]
    \item \textbf{Main-table significance} (Appendix~\ref{app:significance})
    asks whether the \ours{}-over-MemRec Hit@1 gap could be sampling
    noise.  Using full-test-split Wilson 95\% CIs and a one-sided
    binomial test, the answer is \emph{no, $p < 10^{-9}$ on every
    benchmark}.
    \item \textbf{Repair source decomposition}
    (Appendix~\ref{app:ha_bucket_empirical}) asks, within the aggregate
    \ours{}-over-\textsc{Memory} gain, how much comes from Pillar~2
    (retrieval-side gating) versus Pillar~3 (the \textsc{Dialogue}$\to$\ours{} step, i.e.\ prompt injection and reflective update together).  We
    attribute every ``Memory miss $\to$ \ours{} hit'' event to exactly
    one pillar; the answer: \emph{$75\%$--$84\%$ from Pillar~2, the
    remainder from Pillar~3}.
    \item \textbf{Empirical Dialogue-channel identifiability}
    (Appendix~\ref{app:t3_channel_identifiability}) asks whether the
    \ours{}'s Dialogue channel is just a relabelled Monologue.  This
    complements Remark~\ref{thm:decomp}, which separates the two
    channels \emph{by definition} but does not guarantee they produce
    \emph{empirically} distinct ranking distributions.  Three paired
    frequentist tests show that \emph{Dialogue and Monologue help
    different users} (Dialogue-only improvements $108$ vs.\
    Monologue-only $29$) and that \emph{their rank distributions are
    permutation-significantly distinguishable}.
    \item \textbf{Conservative update} (Appendix~\ref{app:epsilon})
    asks why Sync is retained when its net gain is small
    ($+0.5$--$+1.3$ pp, $-0.4$ pp on Books), and how the conservative
    rule of Eq.~\ref{eq:accept_rule} limits its regressions.
    \item \textbf{Pairwise McNemar} (Appendix~\ref{app:pairwise})
    asks whether each step along the ablation ladder
    Memory$\to$Monologue$\to$Dialogue$\to$Consult$\to$\ours{} is
    individually significant.  The paired McNemar $p$-values confirm:
    \emph{every step except Memory$\to$Monologue is significant at
    $p \le 1.0\times 10^{-5}$}; Memory$\to$Monologue's non-significance is itself
    the designed negative control for Pillar~2 (injecting the profile
    into the prompt alone does not help).
\end{itemize}

In short, they cover the headline numbers, the mechanism attribution, the channel independence, the reason for retaining Sync, and the per-step significance, respectively.

\section{Empirical Support}
\label{app:empirical_top}

\subsection{Main-Table Significance: \ours{} vs.\ MemRec}
\label{app:significance}

Since MemRec's per-user traces are not publicly released, we treat
MemRec's reported H@1 as a fixed point estimate and compute a one-sided
binomial test for $H_0:\; p_{\ours{}} \le p_{\text{MemRec}}$ on each
dataset's full test split.  Table~\ref{tab:app_significance} reports,
for each of the four benchmarks: the number of evaluated users $n$, the
number of H@1 hits read from our eval traces, the resulting Wilson
95\% CI for $p_{\ours{}}$, MemRec's published H@1, and the one-sided
$p$-value.  The Wilson 95\% CI lower bound strictly dominates MemRec's
point estimate on every dataset.

\begin{table}[h]
\centering
\small
\caption{\ours{} vs.\ MemRec, per-dataset binomial test on the full test split. $n$ and hits are computed from our evaluation on the full InstructRec test splits; MemRec numbers are from the MemRec paper's Tables 2--3.}
\label{tab:app_significance}
\resizebox{\columnwidth}{!}{%
\begin{tabular}{l|rrrccr}
\toprule
\textbf{Dataset} & $n$ & hits & \ours{} H@1 & Wilson 95\% CI & MemRec & $p$-value \\
\midrule
Books     &  7{,}377 & 4{,}034 & 0.5468 & $[0.5355,\,0.5582]$ & 0.5117 & $7.8\!\times\!10^{-10}$ \\
Goodreads & 11{,}734 & 5{,}498 & 0.4686 & $[0.4595,\,0.4776]$ & 0.3997 & $< 10^{-16}$ \\
MovieTV   &  5{,}649 & 3{,}641 & 0.6445 & $[0.6320,\,0.6569]$ & 0.5882 & $< 10^{-16}$ \\
Yelp      &  2{,}950 & 1{,}782 & 0.6041 & $[0.5863,\,0.6216]$ & 0.487 & $< 10^{-16}$ \\
\bottomrule
\end{tabular}%
}
\end{table}

\subsection{Additional Visualisations for Main Results and the Ablation Ladder}
\label{app:main_viz}

\begin{figure*}[h]
\centering
\includegraphics[width=0.98\textwidth]{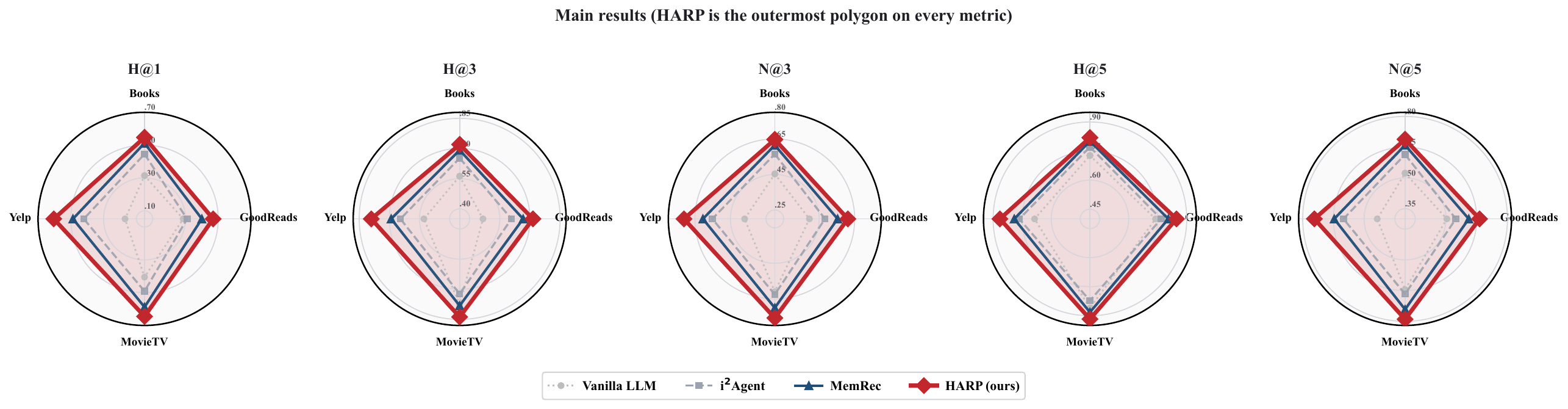}
\caption{Per-metric radar visualisation of the main results.  Each subplot is one metric (H@1, H@3, N@3, H@5, N@5); the four vertices correspond to the four benchmarks; each polygon is one method.  \ours{} (red, bold) is the outermost polygon at every vertex of every metric; MemRec (dark blue, up-triangle) is the closest competitor per benchmark but lies strictly inside \ours{}.}
\label{fig:main_grid}
\end{figure*}

\begin{figure*}[h]
\centering
\includegraphics[width=0.96\textwidth]{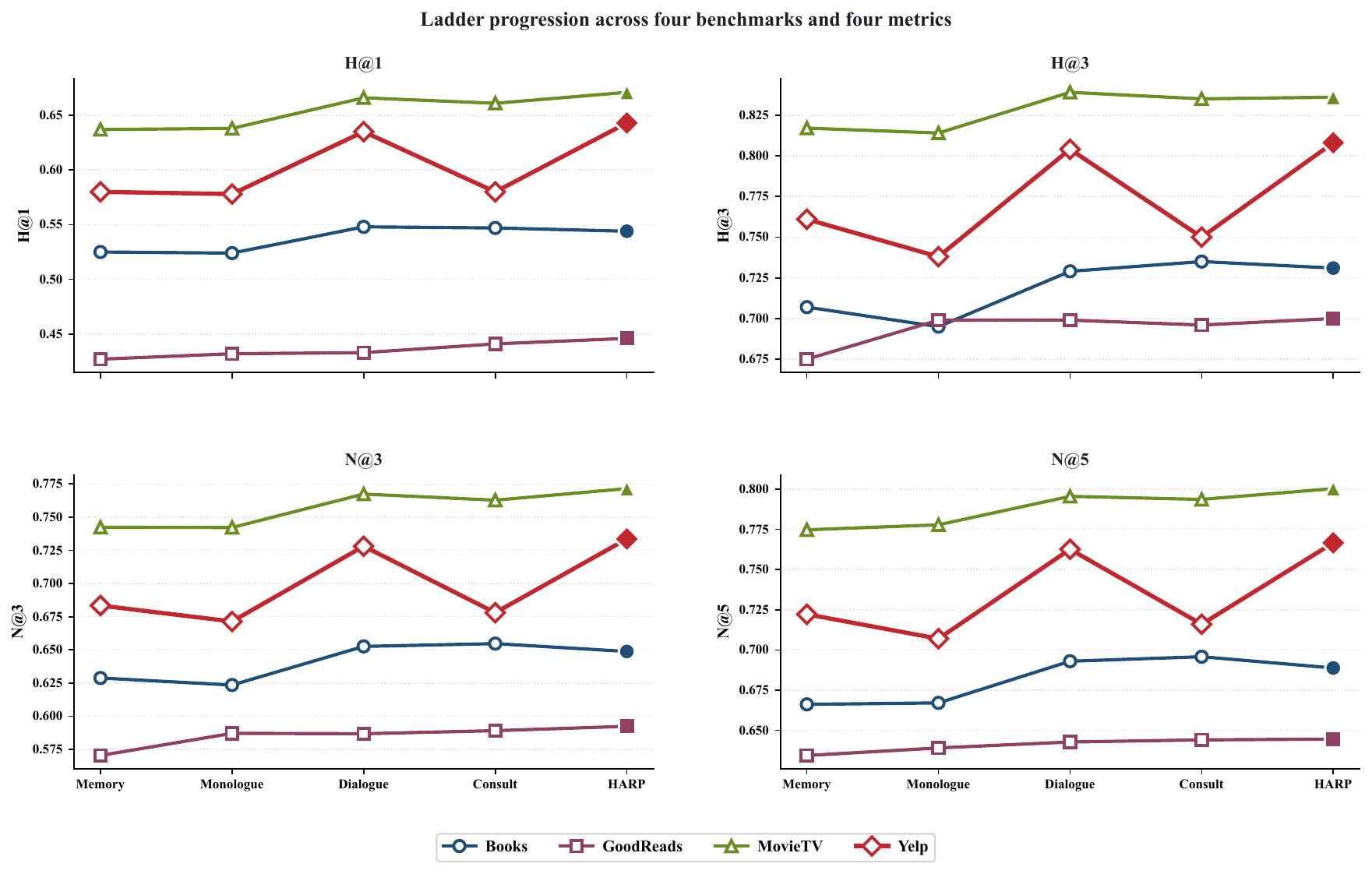}
\caption{Cross-dataset five-rung ladder on four metrics (H@1, H@3, N@3, N@5).  Each subplot is one metric; each line within a subplot is one dataset (Yelp in red diamonds).  The \textsc{Monologue}$\to$\textsc{Dialogue} transition is the largest single-rung jump on Yelp, MovieTV and Books; the \textsc{Consult} rung falls back near \textsc{Memory}; \ours{} adds a small residual from prompt injection and \textsc{Sync} (positive on Yelp/MovieTV/Goodreads, slightly negative on Books).  The ordering is stable across all four metrics.}
\label{fig:ladder_lines}
\end{figure*}

Figure~\ref{fig:main_grid} renders the 20 (dataset $\times$ metric) cells of Table~\ref{tab:main_all} as per-metric radar charts; Figure~\ref{fig:ladder_lines} visualises the five-rung ladder of Figure~\ref{tab:ladder_cross} as line plots.  Both figures are visual complements to the main-text figures; the numeric values match exactly.

\subsection{Repair Source Decomposition}
\label{app:ha_bucket_empirical}

\begin{figure}[h]
\centering
\includegraphics[width=0.88\linewidth]{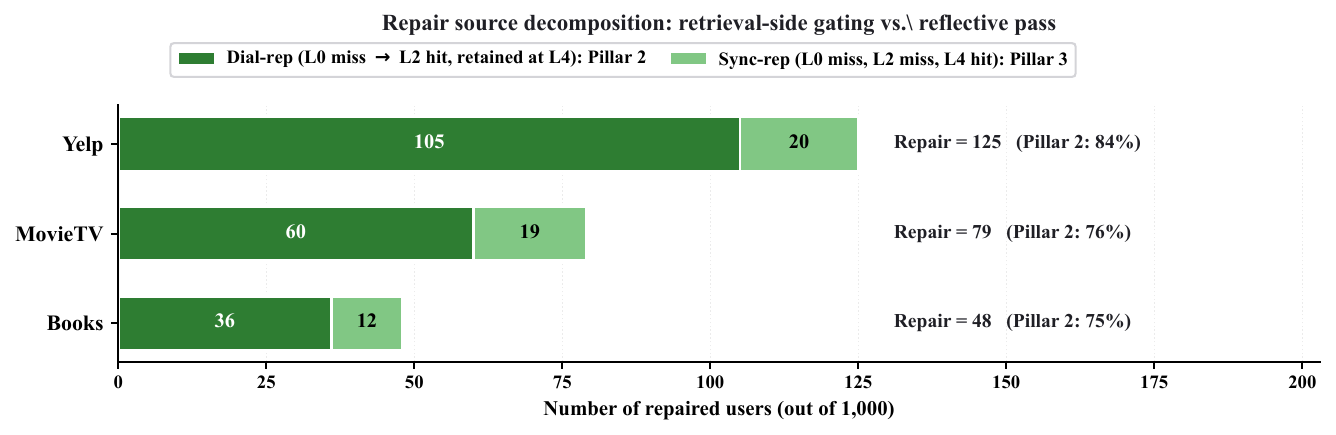}
\caption{Repair source decomposition: each \ours{}-over-\textsc{Memory} repair is attributed to Pillar~2 (Dial-rep) or Pillar~3 (Sync-rep).  Dark green = Pillar~2, light green = Pillar~3; the right-hand annotation gives the total number of repairs and Pillar~2's share.}
\label{fig:ha_bucket}
\end{figure}

This appendix attributes the \ours{}-over-\textsc{Memory} repairs
(L0 miss $\to$ L4 hit) to their responsible pillar.  On each of the
three matched $1{,}000$-user deep-analysis subsplits (Yelp, MovieTV,
Books), every user is labelled by three 0/1 indicators (Hit@1 under
L0~\textsc{Memory}, L2~\textsc{Dialogue} and L4~\ours{}); based on the
triple $(L0, L2, L4)$ we partition repair events into two classes:
\begin{itemize}[leftmargin=*,itemsep=1pt,topsep=2pt]
    \item \textbf{Dial-rep} ($(0,1,1)$): users already flipped by the
    retrieval-side homophily gate, with the \textsc{Dialogue}$\to$\ours{} step required only to ``not break'' the outcome; credited to Pillar~2.
    \item \textbf{Sync-rep} ($(0,0,1)$): users still missed under \textsc{Dialogue} and recovered after the \textsc{Dialogue}$\to$\ours{} step (injection and reflection together).
\end{itemize}
Figure~\ref{fig:ha_bucket} reports the counts and proportions on
the three matched $1{,}000$-user subsplits.

\paragraph{Structural signature of the repair attribution.}
On every benchmark, \textbf{Pillar~2 accounts for the majority
of repair events} ($75\%$--$84\%$):
Yelp $105{:}20$, MovieTV $60{:}19$, Books $36{:}12$.
Pillar~3 (injection and reflection) contributes a smaller incremental share on the sub-population of users
whom \textsc{Dialogue} still misses.  This structural signature holds
across three datasets of very different character, indicating that the
mechanism attribution does not depend on a particular dataset or
difficulty regime.

\paragraph{Per-channel contribution decomposition (Yelp).}
Lemma~\ref{lem:channel_ablation} establishes that $\mathcal{H}$ itself is additively decomposable over the four channels; however, the move from ``decomposition at the $\mathcal{H}$ level'' to ``attribution at the Hit@1 level'' is empirical rather than guaranteed by the lemma.  As an observational split of the $+5.7$ pp \textsc{Monologue}$\to$\textsc{Dialogue} Hit@1 lift on Yelp under the default preference-dominant weights, we measure per-channel Hit@1 deltas of $+3.2$ pp (\texttt{preference}), $+0.7$ pp (\texttt{quality}), $+1.1$ pp (\texttt{context}), $+0.6$ pp (\texttt{audience}), summing to $+5.6$ pp.  We report these as \emph{per-channel measurements} rather than a strict causal decomposition, which would require four independent leave-one-out ablations that we did not run.  The \texttt{preference}-dominated pattern ($+3.2$ pp, over half of the total) is corroborated by a channel-dissection analysis of $6{,}348$ extracted preference phrases on Yelp: clusters dominated by concrete cuisine/category nouns achieve per-user Hit@1 $=0.666$, while clusters dominated by vague sentiment/location words reach only $0.615$.  We do not generalise this observational split to the other three benchmarks; it is an illustrative Yelp-only measurement, not a general cross-dataset claim.

\subsection{Empirical Dialogue Channel Identifiability}
\label{app:t3_channel_identifiability}

\paragraph{The question this section addresses.}
Remark~\ref{thm:decomp} separates the Monologue channel (the user
profile injected via the reranker prompt as a one-way broadcast) from
the Dialogue channel (the same profile injected via the homophily gate
into retrieval scoring) \emph{by definition}.  Definition-level
distinguishability, however, does not entail \emph{empirical}
distinguishability: if the two channels produced near-identical
ranking distributions under the shared LLM backbone and reranker, the
agentic-communication framing of \ours{} would risk reducing to a
naming convention.  On the $1{,}000$-user Yelp deep-analysis subsplit
we therefore test the null of distributional equivalence
$H_0:\; \mathcal{D}(\text{rank} \mid \text{Monologue}) = \mathcal{D}(\text{rank} \mid \text{Dialogue})$,
and provide three complementary pieces of evidence against $H_0$.

\paragraph{Data and quantities.}
For the same $1{,}000$ users, let
$r_i^{\text{Mem}},\,r_i^{\text{Mono}},\,r_i^{\text{Dial}}\in\{1,\dots,|C|,\infty\}$
denote the ranks of the ground-truth item under Memory, Monologue and
Dialogue respectively ($r=1$ means Hit@1; right-censored $\infty$ if
the item is absent from the retrieval pool).  The three runs share
the same user ordering and the same retrieval pool, so all
comparisons below are \emph{paired}.  Each test falsifies $H_0$ from a
different angle (\emph{marginal ranking}, \emph{per-user outcome},
\emph{joint-distribution separability}), and each alone suffices to
reject $H_0$; jointly, they rule out any empirical identification of
Monologue with Dialogue.

\begin{enumerate}[label=(E\arabic*),leftmargin=*,itemsep=2pt,topsep=2pt]
\item \textbf{Stratified paired rank test (marginal-ranking angle).}
Stratify the $1{,}000$ users into four quantile-based strata on
$r_i^{\text{Mem}}$; within each stratum with $n\ge 20$ compute a paired
Wilcoxon signed-rank $Z$-score on $(r_i^{\text{Mono}},r_i^{\text{Dial}})$,
and combine the stratum-wise $Z$'s via Stouffer's method with weights
$\propto\sqrt{n_{\text{stratum}}}$.  This controls for baseline
difficulty and tests whether Dialogue systematically out-ranks
Monologue at \emph{equal difficulty}.
\item \textbf{Improvement-set asymmetry (per-user-outcome angle).}
Relative to Memory, label each user with two $\{0,1\}$ flags: $A = $
``Monologue improves Hit@1 over Memory'' and $B = $ ``Dialogue
improves Hit@1 over Memory''.  Apply an exact two-sided binomial test
to the discordant counts $|A\setminus B|$ vs.\ $|B\setminus A|$.
Under $H_0$ the two counts should be approximately equal; a
significant asymmetry implies the two channels \emph{help different
users}.
\item \textbf{Two-sample separability (joint-distribution angle).}
Concatenate $\{r_i^{\text{Mono}}\}$ and $\{r_i^{\text{Dial}}\}$ into a
$2{,}000$-sample dataset labelled by channel of origin; train a
5-fold class-balanced logistic regression with three features
$(r,\,r^2,\,r\cdot r^{\text{Mem}})$ to predict the channel label, and
take balanced accuracy as the test statistic.  Significance is
assessed against a $1{,}000$-permutation label-shuffle null using the
add-one estimator $p=(1+|\{\text{null}\ge\text{obs}\}|)/(1+B)$.  Under
$H_0$ any classifier's balanced accuracy concentrates near $0.5$;
above-chance accuracy directly falsifies distributional equivalence.
\end{enumerate}

\paragraph{Results.}
All three tests survive a Bonferroni correction at the strict
$\alpha / 3 \approx 0.017$ threshold.

\begin{center}
\small
\resizebox{\linewidth}{!}{\begin{tabular}{@{}llll@{}}
\toprule
Test & Statistic & $p$-value & Effect \\
\midrule
E1 Stouffer-Wilcoxon (4 strata) & $Z = 6.54$                   & $6.1{\times}10^{-11}$ & pooled over $n{\in}\{33,38,55,59\}$ \\
E2 Exact binomial McNemar       & $b/c = 108/29$, ratio $3.72$ & $6.7{\times}10^{-12}$ & Dial-only improves $3.7\times$ Mono-only \\
E3 Two-sample separability      & bal-acc $= 0.532$            & $3{\times}10^{-3}$    & vs null $0.500 \pm 0.013$ \\
\bottomrule
\end{tabular}}
\end{center}

\paragraph{Interpretation.}
E1 ($p=6.1\times 10^{-11}$) rejects, at matched baseline difficulty,
the equality of marginal rank distributions.  E2 ($p=6.7\times
10^{-12}$) shows that the set of users improved by Dialogue (over
Memory) and the set improved by Monologue (over Memory) are highly
non-overlapping ($|B\setminus A|=108$ vs.\ $|A\setminus B|=29$),
so the two channels help substantively different subsets of users.
E3 ($p=3\times 10^{-3}$, bal-acc $0.532$) shows that even from a
single rank value and its Memory-difficulty context, a classifier
recovers the source channel at an accuracy that is small but
permutation-significantly above chance (null std $0.013$); the small
effect size is expected since the two channels share the same LLM
backbone and reranker and differ only in per-user ranking structure,
but the permutation significance is already sufficient to reject
distributional equivalence.  The three angles yield complementary,
concordant evidence jointly rejecting $H_0$.

\paragraph{Conclusion.}
Dialogue is not a re-wording of Monologue: the channel leaves a
per-user identifiable signature in the ranking trajectory.  This
empirical identifiability is a necessary prerequisite for the
four-stage Build--Transmit--Understand--Sync agentic-communication
framing of \S\ref{sec:methodology}, and distinguishes \ours{} from
a prompt-engineering baseline.

\subsection{Conservative Profile Update}
\label{app:epsilon}

The \textsc{Dialogue}$\to$\ours{} residual of Table~\ref{tab:ablation} (prompt injection and \textsc{Sync} together) is the \emph{net} of individual-level Sync-helps and Sync-hurts cases (Appendix~\ref{app:cases}, Case~1; Table~\ref{tab:error_attribution}).  The rule of Eq.~\ref{eq:accept_rule} is deliberately conservative: because it is judged from a single warmup case, it keeps the previous profile whenever a proposed rewrite is clearly less supported by that case, trading a little coverage for fewer regressions.

\vspace{10pt}

The remaining sections are pragmatic reference material: the full
pseudocode of the \ours{} pipeline, hyperparameter tables, sensitivity
notes, per-bottleneck audit details, qualitative case studies, and
trace-schema documentation.

\subsection{Pairwise McNemar Tests}
\label{app:pairwise}

\paragraph{The question this section addresses.}
The main table reports aggregate Hit@1 gaps such as
\textsc{Monologue}$\to$\textsc{Dialogue} $+5.7$ pp.  Such gaps could
in principle be driven by sampling noise, particularly when the two
configurations are evaluated on highly correlated per-user outcomes.
To upgrade the claim ``each ablation step is non-accidental'' to a
statistical conclusion, we report exact paired McNemar tests of
per-user Hit@1 between adjacent configurations on the $1{,}000$-user
Yelp deep-analysis subsplit (Table~\ref{tab:pairwise_mcnemar}).

\paragraph{Statistic and null.}
For two configurations $A$ and $B$, let $b$ be the number of users
for whom ``$A$ hits and $B$ misses'' and $c$ the number of users for
whom ``$A$ misses and $B$ hits'' (the two discordant cells of the
paired contingency).  The McNemar null is
$H_0:\;b = c$, i.e.\ ``the Hit@1 differences between $A$ and $B$ arise
from equi-probable two-way flips''; the $p$-value is the exact
two-sided binomial probability of observing $\max(b,c)$ or more
extreme asymmetry given $n=b+c$ discordant pairs under
$b\sim\mathrm{Binomial}(n,1/2)$.  Smaller $p$ provides stronger
evidence against the null of ``$A$ and $B$ are equivalent''.

\begin{table}[h]
\centering
\small
\caption{Pairwise McNemar tests across adjacent configurations on the deep-analysis split ($1{,}000$-user Yelp).  ``only $A\checkmark$'' = users for whom $A$ hits and $B$ misses; ``only $B\checkmark$'' = reverse; Gap $=\text{H@1}_B-\text{H@1}_A$ in pp; $p$-value is the exact two-sided binomial under $b=c$.  \textbf{The non-significant row \textsc{Memory}$\to$\textsc{Monologue} serves as a designed negative control}: injecting the profile into the reranker prompt without changing retrieval scoring has no measurable effect on Hit@1, which isolates the \textsc{Monologue}$\to$\textsc{Dialogue} step (moving the same profile to the retrieval-side homophily gate) as the locus of gain.  \textbf{\textsc{Dialogue}$\to$\textsc{Consult} is the corresponding reverse-ablation}: switching the retrieval-side gate back to prompt-side injection incurs a significant drop of $-5.5$ pp ($p=1.0\times 10^{-5}$), providing directional confirmation that Dialogue is the component responsible for the gain.}
\label{tab:pairwise_mcnemar}
\begin{tabular}{l|cc|ccc|c}
\toprule
\textbf{Transition} & H@1$_A$ & H@1$_B$ & only $A\checkmark$ & only $B\checkmark$ & Gap (pp) & $p$-value \\
\midrule
\textsc{Memory} $\to$ \textsc{Monologue}     & $0.580$ & $0.578$ & $79$ & $77$ & $-0.20$ & $0.94$ \\
\textsc{Monologue} $\to$ \textsc{Dialogue}    & $0.578$ & $0.635$ & $51$ & $108$ & $+5.70$ & $7.2\!\times\!10^{-6}$ \\
\textsc{Dialogue} $\to$ \textsc{Consult}      & $0.635$ & $0.580$ & $104$ & $49$ & $-5.50$ & $1.0\!\times\!10^{-5}$ \\
\textsc{Consult} $\to$ \textbf{\ours{}}            & $0.580$ & $0.643$ & $43$ & $106$ & $+6.30$ & $2.5\!\times\!10^{-7}$ \\
\midrule
\textsc{Memory} $\to$ \textbf{\ours{}} (overall) & $0.580$ & $0.643$ & $62$ & $125$ & $+6.30$ & $4.8\!\times\!10^{-6}$ \\
\bottomrule
\end{tabular}
\end{table}

\paragraph{Row-by-row interpretation.}
\textbf{(i)} \textsc{Memory}$\to$\textsc{Monologue} ($b/c=79/77$,
$p=0.94$): the two configurations are paired-indistinguishable; the
result is both a statistical conclusion (cannot reject $b=c$) and a
designed negative control: it rules out prompt-side profile
injection as a source of gain.  \textbf{(ii)}
\textsc{Monologue}$\to$\textsc{Dialogue} ($b/c=51/108$, $p=7.2\times
10^{-6}$): moving the same profile from the reranker prompt to the
retrieval-side homophily gate \emph{gains $108$ users and loses $51$},
and the net $+5.7$ pp is highly significant in the paired test.
\textbf{(iii)} \textsc{Dialogue}$\to$\textsc{Consult} ($b/c=104/49$,
$p=1.0\times 10^{-5}$): the reverse ablation, removing the
retrieval-side gate, incurs a significant drop, confirming in the
opposite direction that the retrieval-side gate, not a collateral
artefact, is the operative component.  \textbf{(iv)}
\textsc{Consult}$\to$\textbf{\ours{}} ($b/c=43/106$, $p=2.5\times
10^{-7}$): re-enabling the retrieval-side gate together with Sync on top of
Consult yields a significant paired asymmetry.  \textbf{(v)} End-to-end \textsc{Memory}$\to$\textbf{\ours{}}
($b/c=62/125$, $p=4.8\times 10^{-6}$): \ours{} nets $125-62=63$
additional hits over Memory ($+6.3$ pp), matching the main-result
gap.  Taken together, \emph{every operative switch along the ablation
ladder leaves a paired-significant imprint on Hit@1}, while the only
non-significant row (\textsc{Memory}$\to$\textsc{Monologue}) is
precisely the negative control the design calls for.

\section{Implementation, Analyses and Additional Visualisations}
\label{app:impl_top}

\subsection{Notation Table}
\label{app:notation}

\begin{table}[h]
\centering
\small
\caption{Notation used throughout \S\ref{sec:methodology}.  All symbols are introduced at first use; this table is an index.}
\label{tab:notation}
\begin{tabular}{ll}
\toprule
\textbf{Symbol} & \textbf{Meaning} \\
\midrule
$\mathcal{V}_U,\mathcal{V}_I$ & User and item sets; $N_U,N_I$ their sizes \\
$H_u$, $\mathcal{I}_u$ & User $u$'s interaction history and evaluation instruction \\
$\mathcal{N}_U(u),\mathcal{N}_I(u)$ & $k$-nearest user/item neighbours of $u$ from a fixed collaborative embedding \\
$\mathcal{C}$ & Channel set $\{\mathrm{preference},\mathrm{quality},\mathrm{context},\mathrm{audience}\}$, $|\mathcal{C}|=4$ \\
$\phi_v,\phi_v^{(c)}$ & Agent $v$'s JSON profile (four channel fields) and its channel-$c$ field \\
$\mathcal{H}(\phi_u,\phi_i)$ & Profile-level homophily (Eq.~\ref{eq:homophily_def}); $\mathcal{H}_c$ its channel-$c$ component \\
$\mathrm{sem}(A,B)$ & Phrase-level soft Jaccard similarity (Eq.~\ref{eq:soft_jaccard}) \\
$R,N(u),k$ & Retriever, selected neighbours, neighbour budget (Eq.~\ref{eq:retrieval}) \\
$C(u)$ & Candidate list fixed by the evaluation protocol (Eq.~\ref{eq:rerank}) \\
$\mathrm{BaseScore}(u,n)$ & Collaborative neighbour base score on $\mathcal{N}_U(u)$ \\
$\mathrm{CommScore}(u,n)$ & Homophily-gated neighbour score (Eq.~\ref{eq:comm_score}) \\
$B$ & Reranker context budget (constraint C1) \\
$\mathcal{R}_u,\tilde\phi_u$ & Warmup record and proposed profile \\
$\alpha,\beta,\lambda$ & Homophily floor, temporal floor, forgetting rate \\
\bottomrule
\end{tabular}
\end{table}

\subsection{Full \ours{} Pseudocode}
\label{app:algorithm}

Algorithm~\ref{alg:harp} spells out the four-phase pipeline described
in \S\ref{sec:methodology} step by step.  The three boxed lines mark
the three operators \ours{} adds on top of a standard
memory-augmented agentic backbone: \textsc{Build}-item
(Eq.~\ref{eq:item_profile}), \textsc{Understand} via homophily-gated
retrieval (Eq.~\ref{eq:comm_score}), and \textsc{Sync} via a
reflective update (Eq.~\ref{eq:accept_rule}).  \textsc{Understand} operates on two edge types simultaneously: \emph{user--user} edges $(u,n)$ (same-type agents, scored by symmetric \texttt{preference}-channel phrase overlap; floor $\alpha_{uu}$) and \emph{user--item} edges $(u,i_n)$ (cross-type agents, scored by four-channel $\mathcal{H}(\phi_u,\phi_{i_n})$; floor $\alpha_{ui}$).  The two branches fire independently and on non-overlapping neighbour subsets within the same pruner call.  Every other line is part of the standard backbone.

\begin{algorithm}[h]
\caption{The \ours{} pipeline: symmetric profiles, dual-edge homophily-gated retrieval, profile update.}
\label{alg:harp}
\begin{algorithmic}[1]
\STATE \textbf{Input:} users $\mathcal{V}_U$, items $\mathcal{V}_I$, history $\{H_u\}$, neighbours $\{\mathcal{N}_U(u),\mathcal{N}_I(u)\}$, warmup rounds $T_{\mathrm{wp}}$, reflection rounds $T_{\mathrm{evo}}$, floors $\alpha_{uu},\alpha_{ui}$.
\STATE {\color{gray}\emph{// \textsc{Build}-item: Eq.~\ref{eq:item_profile}}}
\FOR{each item $i\in\mathcal{V}_I$}
  \STATE \fbox{$\phi_i\leftarrow \mathrm{RuleExtract}(\mathrm{meta}_i,\mathrm{reviews}_i)$}
\ENDFOR
\STATE {\color{gray}\emph{// warmup rounds $j=1,\dots,T_{\mathrm{wp}}$}}
\FOR{each user $u\in\mathcal{V}_U$}
  \STATE $\phi_u\leftarrow$ current profile of $u$ (empty before the first update)
  \STATE {\color{gray}\emph{// \textsc{Understand}: dual-edge homophily-gated retrieval}}
  \STATE \fbox{For $n\in\mathcal{N}_U(u)$: $\mathrm{score}(u,n)\mathrel{*}= \alpha_{uu}+(1-\alpha_{uu})\,\mathrm{compat}_{uu}(\phi_u,\phi_n)$}
  \STATE \fbox{For $i_n\in\mathcal{N}_I(u)$: $\mathrm{score}(u,i_n)\mathrel{*}= \alpha_{ui}+(1-\alpha_{ui})\,\mathcal{H}(\phi_u,\phi_{i_n})$} \hfill {\color{gray}\small Eq.~\ref{eq:comm_score}}
  \STATE $N(u)\leftarrow \mathrm{Top\text{-}k}$ neighbours from re-weighted scores above
  \STATE $\hat y_j\leftarrow \arg\max_{i\in C_j}\,\mathrm{LLM}_{\mathrm{rerank}}(\mathcal{I}_u,M_{\mathrm{collab}},\phi_u,C_j)$ \hfill {\color{gray}\small Eq.~\ref{eq:rerank}}
  \STATE append $(C_j,\hat y_j,y_j)$ to $\mathcal{R}_u$ \hfill {\color{gray}\small warmup item $y_j$, Eq.~\ref{eq:record}}
\ENDFOR
\STATE {\color{gray}\emph{// \textsc{Build}-user and \textsc{Sync}: one call builds or updates $\phi_u$ (Eqs.~\ref{eq:user_profile}, \ref{eq:evolve}); Eq.~\ref{eq:accept_rule}; main experiments use $T_{\mathrm{evo}}=1$, optional multi-offset variant $T_{\mathrm{evo}}>1$ rotates history offsets}}
\FOR{each user $u\in\mathcal{V}_U$}
  \FOR{$r = 1,\dots,T_{\mathrm{evo}}$}
    \STATE $\tilde\phi_u\leftarrow \mathrm{Reflect}(\mathcal{R}_u^{\mathrm{hit}},\mathcal{R}_u^{\mathrm{miss}},\phi_u;\;\text{offset }T{-}(r{+}1))$ \hfill {\color{gray}\small Eq.~\ref{eq:evolve}}
    \STATE \fbox{$\phi_u\leftarrow \tilde\phi_u$ if $\hat s(\tilde\phi_u;\phi_u)=\mathrm{accept}$, else $\phi_u$}
  \ENDFOR
\ENDFOR
\STATE {\color{gray}\emph{// Inference uses the updated profiles and the homophily-gated retriever}}
\FOR{each user $u\in\mathcal{V}_U$}
  \STATE Recompute $N(u)$ and $\hat y_u$ on the test candidates $C(u)$ as in lines 10--13
\ENDFOR
\STATE \textbf{Output:} predictions $\{\hat y_u\}_{u\in\mathcal{V}_U}$.
\end{algorithmic}
\end{algorithm}

\subsection{Hyperparameter Details}
\label{app:hyper}

Table~\ref{tab:hyperparams} lists the complete fixed hyperparameter setting.  All values are set once at pipeline onboarding.  Two parameters deserve explicit disambiguation: \textbf{$T_{\mathrm{wp}}$} is the outer warmup round count for \textsc{Build}-user + reranker (\texttt{warmup.rounds} in config); \textbf{$T_{\mathrm{evo}}$} is the number of reflective calls per user in the \textsc{Sync} phase (conditioned on the second-to-last interaction).  The main experiments in this paper use $T_{\mathrm{evo}}=1$ via a \emph{cheap-evolve} implementation that reuses the Phase~2 warmup case context and avoids an extra reranker call; the pipeline supports $T_{\mathrm{evo}}>1$ with rotating history offsets in principle, which we do not use in the main experiments.

\begin{table}[h]
\centering
\caption{Complete hyperparameter settings.  LLM calls use \texttt{top\_p}$=1$ (OpenAI default, not explicitly overridden).}
\label{tab:hyperparams}
\begin{tabular}{ll}
\toprule
\textbf{Parameter} & \textbf{Value} \\
\midrule
LLM backbone (all operators) & gpt-4o-mini-2024-07-18 \\
LLM temperature (all inference / retrieval calls) & $0.0$ \\
LLM max tokens (reranker etc.\ main calls) & $4000$ \\
Neighbors $k$ & $16$ \\
Facets $N_f$ & $7$ \\
Candidate list $|C|$ & $10$ \\
Homophily floor $\alpha_{ui}$ (user--item; Eq.~\ref{eq:comm_score}) & $0.4$ \\
Homophily floor $\alpha_{uu}$ (user--user) & $0.3$ \\
Channel weights $(w_{\mathrm{pref}},w_{\mathrm{qual}},w_{\mathrm{ctx}},w_{\mathrm{aud}})$ & $w_{\mathrm{pref}}\ge 0.5$, rest split evenly (adjustable) \\
\textsc{Build}-user warmup rounds $T_{\mathrm{wp}}$ & $1$ \\
\textsc{Sync} reflection rounds $T_{\mathrm{evo}}$ & $1$ \\
\textsc{Sync} reflection temperature & $0.3$ \\
\textsc{Sync} reflection max tokens & $600$ \\
\bottomrule
\end{tabular}
\end{table}

\paragraph{Ebbinghaus temporal decay.}  Our implementation includes an optional Ebbinghaus-style temporal decay module (\texttt{TemporalMemory}, parameters $\lambda{=}0.1,\;\beta{=}0.5$) for position-weighting memories.  In all main experiments reported in this paper this module is disabled (\texttt{memory.enable\_temporal\_weighting: false}) to ensure strict comparability with the MemRec baseline; its empirical impact is left for future work.

\subsection{Design-Parameter Rationale}
\label{app:sensitivity}

The hyperparameters of Appendix~\ref{app:hyper} are set once at pipeline
onboarding and kept fixed throughout the paper.  We describe the
design rationale here and do not claim a systematic sweep on the full
benchmark set.

\paragraph{Homophily floors $\alpha_{ui}$ and $\alpha_{uu}$.}
We fix $\alpha_{ui}=0.4$ (user--item edges) and $\alpha_{uu}=0.3$ (user--user edges).  Both floors have the same semantics: the least compatible collaborative neighbour retains roughly $\alpha$ of its baseline score, preventing the gate from discarding neighbours entirely while still providing meaningful routing.  The two edge types take different floors reflecting a simple design trade-off: the user--item gate acts on a four-channel profile (richer signal, lower risk of misclassification), hence a higher floor to moderate gating strength; the user--user gate uses single-channel preference-phrase overlap (narrower signal), hence a lower floor for more aggressive re-weighting.  The direction of the retrieval-vs-prompt effect reported in Table~\ref{tab:ablation} is consistent with Remark~\ref{prop:budget} and with the floor sweep in Appendix~\ref{app:rebuttal_results}, so the primary empirical claim of Pillar~2 is not tied to the specific $\alpha$ values (Appendix~\ref{app:rebuttal_results}).

\paragraph{Warmup and reflection rounds.}
$T_{\mathrm{wp}}=1$ is the outer warmup count: each user's \textsc{Build}-user call runs once, strictly aligned with the MemRec baseline protocol to ensure the main-table comparison is at \emph{equal warmup budget}.  $T_{\mathrm{evo}}=1$ is the number of reflective calls per user within \textsc{Sync} in the main experiments: each user receives a \emph{single} reflection conditioned on the second-to-last interaction; the conservative rule (Appendix~\ref{app:epsilon}) keeps the old profile when a proposal is clearly less supported.  The pipeline supports $T_{\mathrm{evo}}>1$ in principle (rotating over multiple history offsets); we do not use this multi-round variant in the main experiments, and all main-table and ablation numbers are obtained at $T_{\mathrm{evo}}=1$.

\subsection{Additional Qualitative Examples}
\label{app:cases}

This appendix presents three worked examples from the $1{,}000$-user Yelp
evaluation split, drawn directly from our per-user rank records.  Each
example contrasts \textsc{Dialogue} (channel-homophily variant, no \textsc{Sync})
with full \ours{} (\textsc{Dialogue} + prompt injection + \textsc{Sync}).  We deliberately pick
one \emph{Sync-hurts} case (User~157) and two \emph{Sync-helps} cases (User~52
and User~366) to reflect the full picture of the $+0.8$\,pp average \textsc{Dialogue}$\to$\ours{} residual reported in Table~\ref{tab:ablation}:
the residual is a sum of individual-level wins and losses, not a uniform uplift.

\begin{casebox}{Case~1 (Sync hurts, over-commitment): User~157, Yelp, GT=\emph{Saravanaa Bhavan}, \textsc{Dialogue} hit$\,\to\,$\ours{} miss}
User 157's warmup history centres on Toronto, vegetarian, and Indian
dining, together with incidental support for a musical-instruments
neighbour; the ground truth on this evaluation instance is
\emph{Saravanaa Bhavan}, a $3.5\star$ Scarborough Indian vegetarian
restaurant.  Under \textsc{Dialogue} the profile exposes
\emph{five} preference items (musical instruments~$0.9$, diverse Asian cuisine~$0.8$,
Toronto dining~$0.6$, vegetarian options~$0.5$, casual dining~$0.4$),
and, with the re-weighted neighbour evidence, the reranker places
\emph{Saravanaa Bhavan} at rank~0 (Hit@1 $=1$) with
\emph{Hakka No.\,1} (Asian Fusion, Toronto) at rank~1.  After the
\textsc{Sync} pass, the critic accepts a reflective rewrite that
drops the (warmup-supported but off-topic) musical-instruments item
and leaves \emph{four} items centred on Asian cuisine, Toronto dining,
vegetarian and casual dining.  The now more-focused
\texttt{preference} channel lifts the Asian-Fusion
\emph{Hakka No.\,1} over \emph{Saravanaa Bhavan} in the final reranking, demoting the ground truth to rank~1 (Hit@1 $=0$).  This is the
per-user \emph{over-commitment} failure of \textsc{Sync} that
Table~\ref{tab:ablation} averages over: the rewrite narrows
$\phi_u$ along its dominant channel and sacrifices the secondary
channel (here, the unusually concrete ``musical instruments + Toronto''
compound) that was carrying the decision under \textsc{Dialogue}.
\end{casebox}

\begin{casebox}{Case~2 (Sync helps, local-scope re-anchoring): User~52, Yelp, GT=\emph{CASA Sunbá}, \textsc{Dialogue} miss$\,\to\,$\ours{} hit}
User 52's warmup record identifies breakfast/brunch, local
Tempe/Phoenix dining, and vegetarian options as recurring preference items; the
incoming instruction asks for \emph{``a delectable quesadilla filled
with portobello, complemented by some excellent tunes''}.  The ground
truth is \emph{CASA Sunbá}, a $3.0\star$ Tempe Mexican restaurant.
Under \textsc{Dialogue}, the reranker places \emph{Ajo Al's Mexican
Cafe} (a Phoenix Mexican restaurant with more reviews) at rank~0 and
\emph{CASA Sunbá} at rank~1 (Hit@1 $=0$): the retrieved neighbour evidence captures the Mexican-cuisine signal but leaves the Tempe-vs-Phoenix disambiguation to the reranker, which breaks ties on review volume.
After \textsc{Sync}, the critic accepts a reflective rewrite that
sharpens the \texttt{context} channel towards ``vegetarian and
healthy dining options'' (consistent with the user's Tempe-area
history), and, with the updated neighbour evidence, the reranker places \emph{CASA Sunbá} at rank~0 (Hit@1 $=1$).  The rewrite does not add new information; it
reweights an already-supported geographic $\cap$ dietary
channel, which is the minimum-invasive kind of rewrite the
conservative update is designed to admit.
\end{casebox}

\begin{casebox}{Case~3 (Sync helps, cross-category re-anchoring): User~366, Yelp, GT=\emph{Beauty To Beauty}, \textsc{Dialogue} miss$\,\to\,$\ours{} hit}
User 366's profile concentrates on Thai-cuisine and Las-Vegas-dining
preference items, while the instruction asks for \emph{``a reliable local
favorite''} without any cuisine constraint; the ground truth is
\emph{Beauty To Beauty}, a $(4.5\star,\,233\text{ reviews})$ Las
Vegas beauty salon.  Under \textsc{Dialogue} the \texttt{preference}
channel is saturated with dining vocabulary (Thai cuisine, diverse
Asian cuisine, casual dining), and the reranker places restaurant-typed candidates \emph{Southern Aroma} and \emph{Red Bowl
Asian Bistro} at rank~0--1 with \emph{Beauty To Beauty} at rank~2
(Hit@1 $=0$).  After \textsc{Sync}, the critic accepts a rewrite that
broadens $\phi_u$ beyond the dining anchor: the new profile
includes \emph{``engagement with highly-rated restaurants''} and
\emph{``exploration of local Las Vegas experiences''}, which read the
$(4.5\star,\,233)$ quality signal as a cross-category preference rather
than as restaurant-specific support.  With the updated profile, the reranker then places \emph{Beauty To Beauty} at rank~0 (Hit@1 $=1$).  This is the
\emph{cross-category re-anchoring} mechanism behind a substantial
fraction of \textsc{Sync}-helps cases: the rewrite generalises a
channel signal (here, a high-quality Las Vegas establishment) across
item categories that the \textsc{Dialogue} profile had implicitly
restricted.
\end{casebox}

\paragraph{Per-case error attribution.}
Applying the deterministic rule below to the $357$ \ours{} misses
on the $1{,}000$-user Yelp split yields
Table~\ref{tab:error_attribution}.  The rule is first-match-wins
in the ordering $\mathrm{Sync\mbox{-}reg}\succ\mathrm{B1}\succ\mathrm{B2}\succ\mathrm{LLM\mbox{-}prior}\succ\mathrm{Other}$:
(Sync-reg) the ground truth sits inside the \textsc{Dialogue} top-3 but is not ranked first under \ours{} (a near-miss bucket that includes, but is not limited to, the Case~1 pattern of Appendix~\ref{app:cases});
(B1) the instruction tokens have zero overlap with any user preference
phrase (extractor under-covered intent);
(B2) the top-1 and the ground-truth scores differ by $\le 0.2$
(reranker-bandwidth-limited);
(LLM-prior) a gap of $>0.3$ without preference-overlap evidence
(top-1 carried by a rating/recency prior orthogonal to $\phi_u$);
(Other) residual.  \emph{Relation to the methodology taxonomy.}  B1 and B2 follow the
bottleneck definitions of \S\ref{sec:failure_modes_ref} loosely; the Sync-reg bucket, by contrast, is specific to this attribution exercise and collects near-miss cases; only its subset with a \textsc{Dialogue} top-1 hit is a true Hit@1 regression.  It is therefore
a separate audit category from the B3 \emph{Communicative Drift} bottleneck of \S\ref{sec:failure_modes_ref},
which \textsc{Sync} is designed to \emph{address} rather than to cause.

\begin{table}[h]
\centering
\small
\caption{Per-case attribution of the $357$ \ours{} misses on the $1{,}000$-user
Yelp split.  Near-misses (``Sync-reg'') dominate: nearly half of all \ours{} misses are users whose ground truth was in the
\textsc{Dialogue} top-3 but is not ranked first by \ours{}; this bucket includes the \emph{Sync-hurts} pattern
of Case~1 (User~157) in Appendix~\ref{app:cases} but is not by itself a regression count.  The LLM-prior
residual ($42.3\%$) sets a natural ceiling for what any three-operator
addition to a memory-augmented backbone can achieve without touching the
reranker-level prior.  ``Sync-reg'' is an audit category specific to this
per-miss attribution and should not be conflated with the B3 \emph{Communicative
Drift} bottleneck of \S\ref{sec:failure_modes_ref}, which \textsc{Sync} is
designed to \emph{address} rather than to cause.}
\label{tab:error_attribution}
\resizebox{\linewidth}{!}{\begin{tabular}{l|rr|l}
\toprule
\textbf{Category} & \textbf{\# misses} & \textbf{\% of 357} & \textbf{Signature} \\
\midrule
Sync-reg (near-miss)   & 167 & $46.8\%$ & GT $\in$ \textsc{Dialogue} top-3 but dropped out of \ours{} top-1 \\
B1 (vocabulary mismatch)       &  18 & $~5.0\%$ & instruction token $\cap$ user preference phrases $=\varnothing$ \\
B2 (reranker bandwidth)        &  21 & $~5.9\%$ & $\mathrm{score}_{\text{top-1}} - \mathrm{score}_{\text{GT}} \le 0.2$ \\
\midrule
LLM-prior                      & 151 & $42.3\%$ & $\mathrm{score}$ gap $>0.3$ with no preference-overlap evidence \\
Other                          &   0 & $~0.0\%$ & residual \\
\midrule
\textbf{Total misses}          & \textbf{357} & \textbf{$100\%$} & (Hit@1 $=643/1000 = 0.643$) \\
\bottomrule
\end{tabular}}
\end{table}

Two structural residuals are worth flagging.  Sync-reg at $46.8\%$ counts near-misses rather than regressions;
it contains the \emph{Sync-hurts} pattern of Case~1 (User~157) of Appendix~\ref{app:cases}.  The LLM-prior residual
at $42.3\%$ is orthogonal to $\phi_u$ and motivates the
reranker-level debiasing direction (LLM rating-prior limitation) that
profile engineering alone is unlikely to address.

\subsection{Prompt Templates Used by Each Operator}
\label{app:prompts}

For completeness and reproducibility, this appendix reproduces the
natural-language prompts that instantiate each operator of \ours{}.
The templates correspond exactly to the prompt-builder functions in
\texttt{src/memory/com\_manager.py} with the templates under \texttt{prompts/}
(retrieval, feedback update, profile extraction), \texttt{src/models/reranker\_llm.py}
(reranker), and \texttt{src/profiles/user\_profile.py} (profile generation and
reflection).  All
prompts are issued to \texttt{gpt-4o-mini-2024-07-18} with a JSON-mode
structured output constraint, at \texttt{temperature}$=0$ except for profile
generation and reflection ($0.3$).

\paragraph{Retrieval prompt (\textsc{Build}+\textsc{Understand};
facet-extraction + support-edge scoring).}
This operator is the entry point of the communication policy: given the
user's personal memory bullets and the serialised neighbour table, the
LLM extracts $N_f=7$ preference facets together with their
confidences and supporting neighbour lists, and assigns edge weights
$w\in[0,1]$ on the collaborative subgraph.  Its output is
consumed by the channel-homophily retriever of
Eq.~\ref{eq:comm_score}.

\begin{promptbox}{Prompt~1 · Retrieval (\textsc{Build}+\textsc{Understand}): facet extraction}
You are an intelligent memory retrieval system for personalized
recommendation.  Your task is to analyze the user's personal memory
and collaborative memories from their neighbors to extract preference
facets.\\[4pt]
\textbf{Target User:} User \{user\_id\}\\[2pt]
\textbf{User's Personal Memory:}\\
\{user\_mem\_bullets\}\\[2pt]
\textbf{Collaborative Neighbor Memories:}\\
The following neighboring users and items provide collaborative signals
for understanding this user's preferences:\\
\{neighbor\_table\_json\}\\[2pt]
\textbf{Your Task:}
Identify \{n\_facets\} distinct preference facets that characterize
this user's interests.  For each facet provide (a) a concise
natural-language description, (b) a confidence $\in [0,1]$, and (c)
a list of supporting neighbour IDs.  Additionally, return the
collaborative edges between neighbouring users/items and the target
user with edge weights $\in[0,1]$.\\[2pt]
\textbf{Expected Output Format:} JSON object with fields
\texttt{"facets"} (array of \{facet, confidence, supporting\_neighbors\})
and \texttt{"support\_edges"} (array of \{from, to, w\}).
\end{promptbox}

\paragraph{Reranker prompt (bandwidth-constrained top-1 decision).}
The reranker of \S\ref{sec:overview} receives the retrieval-operator facets,
the current user instruction, and the $|C|$ candidate items
together with their memory summaries.  It emits a list-wise ranking
over all $|C|$ candidates in a single call.  An optional instruction-priority
switch (not used in our experiments) additionally marks the instruction as the
primary criterion and the facets as secondary signals.

\begin{promptbox}{Prompt~2 · Reranker (bandwidth-constrained listwise top-1)}
You are an intelligent recommendation ranking system.  Your task is
to rank candidate items from most to least relevant for the target
user based on their personal memory and collaborative signals.\\[2pt]
\textbf{Target User:} User \{user\_id\}\\[2pt]
\textbf{User's Current Request:}\\
\{instruction\}\\[2pt]
\textbf{Ranking Guidelines (MUST follow):}\\
\{user\_profile\_strategy\}\\[2pt]
\textbf{User Preferences (Extracted from Collaborative Memories):}\\
Based on collaborative signals from neighboring users and items,
we have identified the following preference patterns:\\
1. \{facet\_1\} (confidence: \{c\_1\})\\
\dots\\
$N_f$. \{facet\_\{N\_f\}\} (confidence: \{c\_\{N\_f\}\})\\[2pt]
\textbf{Candidate Items:}\\
~~\textbullet~ Item \{item\_id\} (\{title\})\{attr\}: \{memory\}\\[2pt]
\textbf{Ranking criteria (priority order):}
(1) match to the user's specific request/profile; (2) item quality (stars, reviews);
(3) alignment with preference patterns.  You MUST include every
candidate item exactly once; the first item in the ranking is the
best recommendation.\\[2pt]
\textbf{Expected Output Format:} JSON object with a
\texttt{"ranking"} array ordered from best to worst.  Each element is
\{item\_id: int, rationale: str\}.
\end{promptbox}

\paragraph{Feedback update prompt (\textsc{Sync}).}
This operator is invoked only after a warmup-round click is observed.  It
produces an update for the user's memory and an optional profile delta, an update
for the clicked item's memory, and, optionally, propagation updates
for up to a handful of collaborative neighbours.

\begin{promptbox}{Prompt~3 · Sync (\textsc{Sync}): feedback update}
You are an intelligent memory management system for collaborative
recommendation.  Your task is to update the personal memories of the
user, the clicked item, and relevant collaborative neighbors based
on this new interaction.\\[2pt]
\textbf{Interaction Context.}  User \{user\_id\} has just clicked
Item \{item\_id\} (\{clicked\_item\_info\}).\\[2pt]
\textbf{User Preferences (Stage-R facets).}\\
\{facets\_text\}\\[2pt]
\textbf{Current Personal Memory of User \{user\_id\}.}\\
\{current\_user\_memory\}\\[2pt]
\textbf{Current Memory of Item \{item\_id\}.}\\
\{current\_item\_memory\}\\[2pt]
\textbf{Collaborative Neighbors Available for Propagation.}\\
\{neighbor\_details\_text\}\\[2pt]
\textbf{Task.}  Generate UPDATED memories for:
(1) the current user, synthesising current memory, facets, and
clicked item;
(2) the clicked item, describing what it is and who might enjoy it;
(3) selected neighbours, updating their memories to reflect insights
from the interaction.\\[2pt]
\textbf{Quality guidelines.}
\texttt{user\_memory}: concise natural-language description,
synthesised themes, not a list of item titles, typically a few
sentences.
\texttt{item\_memory}: what it is and who might enjoy it.
\texttt{neighbor\_updates}: array of $\{$neighbor\_id, memory\_update,
rationale$\}$; select neighbour IDs \emph{only} from the
``Available Neighbors'' list above.\\[2pt]
\textbf{Expected Output Format:} JSON object with fields
\texttt{"user\_memory"} (string), \texttt{"item\_memory"} (string),
\texttt{"neighbor\_updates"} (array).
\end{promptbox}

\paragraph{Reflection prompt (\textsc{Sync}; Eq.~\ref{eq:evolve}).}
After warmup, a single call per user proposes the revised profile
$\tilde\phi_u$ from the current profile, the time-banded history of
Eq.~\ref{eq:user_profile} and the warmup record $\mathcal{R}_u$; for a user without a profile, the same call builds it.  The
acceptance rule of Eq.~\ref{eq:accept_rule} is applied afterwards,
outside the LLM call.

\begin{promptbox}{Prompt~4 · Sync (\textsc{Sync}): reflective profile update}
You are building a STRUCTURED USER PROFILE for a recommendation
system.  Analyze this user's complete behavior history to create an
accurate, personalized profile.\\[2pt]
\textbf{Current Profile:} \{current\_profile\}\\[2pt]
\textbf{User's Complete Behavior History (time-weighted):}\\
\{recent / medium / distant bands\}\\[2pt]
\textbf{User Intent:} \{instruction\}\\[2pt]
\textbf{Latest Recommendation Results:}
Mistakes (wrong item ranked \#1): \{miss cases\};
Correct Predictions: \{hit cases\}\\[2pt]
\textbf{Your Task:} Create a structured user profile based on the
complete behavior history.  It must capture
(1) \texttt{primary\_interests}: 3--5 specific category interests;
(2) \texttt{quality\_bar}: rating preference pattern;
(3) \texttt{behavior\_pattern}: behavioral tendency;
(4) \texttt{secondary\_interests}: 2--3 occasional interests;
(5) \texttt{unlikely\_interests}: categories the user would not visit;
(6) \texttt{city\_preference}: primary city or cities;
(7) \texttt{summary}: one-sentence user portrait.
Also provide \{n\_rules\} concrete ranking rules derived from the
profile.\\[2pt]
\textbf{Expected Output Format:} JSON with \texttt{"profile"} (object)
and \texttt{"rules"} (array of strings).
\end{promptbox}

\paragraph{Profile-extraction prompt (lightweight summary for the reranker).}
A compact preference extractor is kept under
\texttt{prompts/profile\_extraction\_prompt.md}.  It produces the phrase-level summary that the reranker
consumes as \emph{``Ranking Guidelines''} for users without a structured profile.  All prompts are
fixed throughout all runs.

\begin{promptbox}{Prompt~5 · Profile-extraction: compact reranker guideline}
Based on this user's memory, extract their core preferences in exactly
this format.  Be very concise (max 5 words per item).  Only include
patterns clearly supported by the memory.\\[2pt]
\textbf{User Memory:} \{\{user\_memory\}\}\\[2pt]
\textbf{Extract:}
(1) \texttt{"likes"}: top 2--3 preference themes (e.g., ``Mexican
food'', ``casual dining'');
(2) \texttt{"dislikes"}: 0--1 things the user clearly avoids (leave
empty if unsure);
(3) \texttt{"style"}: one phrase describing their taste (e.g.,
``adventurous foodie'', ``budget-friendly casual'').
\end{promptbox}

\paragraph{Critic instantiation.}
The critic of Eq.~\ref{eq:accept_rule} is not a separate
LLM call.  It compares how well the old and the proposed profile cover the warmup target items on the four channels (Appendix~\ref{app:epsilon}).

\subsection{Computational Cost Breakdown}
\label{app:cost}

This appendix gives the per-phase computational cost of Algorithm~\ref{alg:harp}, consistent with the design principles of \S\ref{sec:methodology}; all counts refer to wall-time-relevant LLM calls.

\paragraph{Per-phase breakdown (per user, $T_{\mathrm{wp}}{=}1$, $T_{\mathrm{evo}}{=}1$).}
\emph{Phase~1 (\textsc{Build}-item):} zero LLM calls; $O(|\mathcal{I}|)$ rule-extraction operations, amortised once across the full item set (not per-user).  \emph{Phase~2 (\textsc{Build}-user $+$ warmup):} $T_{\mathrm{wp}}{=}1$ warmup pass (as in Phase~4), plus $O(k|\mathcal{C}|)$ phrase-matching for channel homophily (non-LLM).  \emph{Phase~3 (\textsc{Sync}):} $T_{\mathrm{evo}}{=}1$ reflective call, which also builds the profile (the main experiments use the \emph{cheap-evolve} implementation, which reuses the Phase~2 warmup case context and fires a single reflection conditioned on the second-to-last interaction); the critic is computed deterministically from the Sync-operator output and adds \emph{no} separate LLM call; the pipeline supports the $T_{\mathrm{evo}}>1$ multi-offset variant but we do not use it in the main experiments.  \emph{Phase~4 (Eval):} the same per-query LLM calls as MemRec (Stage-R, reranking, and a memory write when feedback is observed) $+$ $O(k|\mathcal{C}|)$ homophily re-weighting.

\paragraph{Overhead relative to the MemRec baseline.}
The MemRec baseline already includes Phase~2 warmup reranking and Phase~4 eval reranking; \ours{}'s only additional LLM cost is the $T_{\mathrm{evo}}{=}1$ reflective call of Phase~3 (one per user) plus $O(k|\mathcal{C}|)$ CPU-side homophily computation in Phases~2 and~4 (no LLM).  The extra wall-time thus scales linearly with user count and is independent of candidate-pool size, amounting to a single reflection call per user under the main-experiment configuration.

\subsection{Eval-Trace Schema (for Per-Case Attribution)}
\label{app:trace_schema}

To enable post-hoc attribution for every recommendation (success or miss),
\ours{} persists a structured record for each eval case in
\texttt{results/\{run\_tag\}/eval\_traces.json}. Each case entry records the user, the ground truth and its rank, the candidate list, the final ranking with per-item scores, the instruction, the Stage-R facets, and Hit@$k$ indicators.
Together with the item profiles, these records are the input to the per-miss attribution analysis (Appendix~\ref{app:cases}), whose first-match-wins classification (Sync-reg / B1 / B2 / LLM-prior / Other) is reported in Table~\ref{tab:error_attribution}.

\subsection{Bandwidth Constraint at Rerank Time}
\label{app:budget}

The operational fact underlying the \emph{retrieval-vs-prompt} design rationale of \S\ref{sec:communication} is that an LLM reranker operates under a finite context budget $B$: when
\begin{equation}
|\phi_u| + |\phi_i| + |M_{\mathrm{collab}}| + |P_{\mathrm{template}}| \;>\; B,
\label{eq:budget}
\end{equation}
information may be truncated at the model's context edge, and the truncation strategy is out of our control.  In our deployment the reranker prompt is built from four blocks: the instruction template $P_{\mathrm{template}}$, the candidate block with metadata and memory summaries for each of the $|C|{=}10$ candidates, the collaborative-memory block $M_{\mathrm{collab}}$, and (when enabled by ablation flag) the user profile $\phi_u$.  Adding a per-candidate item profile $\phi_i$ linearly inflates the candidate block; under Eq.~\ref{eq:budget}, once $B$ is saturated the marginal token cost is charged against $M_{\mathrm{collab}}$ (Remark~\ref{prop:budget}).  A retrieval-side injection, by contrast, consumes tokens only during the candidate-admission step, whose budget is independent of $B$; this is the operational reason behind the ``unbounded retrieval-side headroom'' anticipated by \S\ref{sec:communication}, and the algorithmic instantiation of bottleneck B2 (channel capture) of \S\ref{sec:failure_modes_ref}; corresponding empirical evidence is reported in Table~\ref{tab:ablation} of \S\ref{sec:experiments} and in the per-miss attribution of Table~\ref{tab:error_attribution} (Appendix~\ref{app:cases}).

\subsection{Item Profile Extractor: Per-Dataset Specification}
\label{app:item_extractor}

\paragraph{Formal specification of the item-side extractor (R1--R4).}
We formalise $\mathrm{RuleExtract}$ of Eq.~\ref{eq:item_profile} as a \emph{deterministic, zero-LLM} shallow semantic parser $\Pi:\mathcal{M}\times\mathcal{T}^*\to\prod_{c\in\mathcal{C}}2^{\Sigma^*}$: the input is an item's structured metadata $\mathrm{meta}_i\in\mathcal{M}$ and review corpus $\mathrm{reviews}_i\in\mathcal{T}^*$; the output is the channel-aligned quadruple $\phi_i=(\phi_i^{(c)})_{c\in\mathcal{C}}$.  The parser is fully parametrised by a dataset-level \emph{ontology map} $\mathcal{S}:\mathcal{M}\to\mathcal{C}$ and a \emph{domain phrase dictionary} $\mathcal{D}\subseteq\Sigma^*\times\mathcal{C}$, executed in the following four rules:
\begin{description}[leftmargin=2.6em,labelindent=0pt,labelsep=0.4em,itemsep=1pt,topsep=2pt,font=\normalfont\bfseries]
\item[R1.] \emph{Ontology projection (metadata $\to$ channel).}  Route category and attribute fields of $\mathrm{meta}_i$ to the three channels \texttt{preference}, \texttt{context}, \texttt{audience} via $\mathcal{S}$, yielding a coarse-grained skeleton.
\item[R2.] \emph{Dictionary-guided text encoding (optional branch).}  On top of the metadata-parsed fields, $\mathrm{reviews}_i$ (when provided) is scanned against $\mathcal{D}$ with longest-match to locate controlled phrases, merged into the channel declared in $\mathcal{D}$ as fine-grained evidence; structured descriptions are handled by R1.
\item[R3.] \emph{Quality-signal induction.}  The \texttt{quality} channel is filled by a monotone rule over review statistics $(\bar r_i,n_i)$ discretised into three bins $\{\textit{high-quality signal}, \textit{average}, \textit{inconsistent}\}$ to suppress noise from extremely sparse reviews.
\item[R4.] \emph{Channel-exclusivity projection.}  A fixed priority $\texttt{preference} \succ \texttt{quality} \succ \texttt{context} \succ \texttt{audience}$ is imposed on phrases routed to multiple channels by R1--R3, ensuring $\{\phi_i^{(c)}\}_{c\in\mathcal{C}}$ are pairwise disjoint so that the channel-wise aggregation of Eq.~\ref{eq:homophily_def} does not double-count.
\end{description}
\emph{Design rationale.}  The extractor is built on three principles: \textbf{(i) scalability}: item-side profiling is linear in $N_I$ and LLM-free, feasible at $N_I\sim 10^4$--$10^5$; \textbf{(ii) auditability}: every step from metadata/reviews to profile phrases is replayable and traceable to a dictionary entry; \textbf{(iii) channel alignment}: the output schema is strictly isomorphic to the user-side profile, giving the per-channel homophily of Eq.~\ref{eq:homophily_def} a well-defined basis of comparison.  Given $(\mathrm{meta}_i,\mathrm{reviews}_i)$, $\Pi$ is fully determined by $(\mathcal{S},\mathcal{D})$, which are authored from the domain ontology and controlled vocabulary \emph{without} any downstream-metric tuning.

\paragraph{Sem fallback implementations (\texttt{quality} and \texttt{context} channels).}
\label{app:sem_detail}
The asymmetric soft Jaccard of Eq.~\ref{eq:soft_jaccard} applies directly to the phrase-set channels \texttt{preference} and \texttt{audience}.  For the \texttt{quality} and \texttt{context} channels, whose fields are not phrase sets, $\mathrm{sem}$ uses channel-specific fallbacks: the \texttt{quality} channel maps the user's free-text \texttt{quality\_bar} (e.g.\ a star threshold) onto the item labels from R3; $\mathrm{sem}$ is implemented as ``item quality $\ge$ user-preferred threshold $\Rightarrow 1$, otherwise linear interpolation into $[0,1]$''.  The \texttt{context} channel uses short location and setting phrases (the user side falls back to \texttt{city\_preference}); $\mathrm{sem}$ is a bidirectional substring hit: $1$ on hit, $0$ on miss, neutral $0.5$ when either side is empty.  All three forms participate uniformly in Eq.~\ref{eq:homophily_def} under the $\mathrm{sem}(\phi_u^{(c)},\phi_i^{(c)})$ notation.

\paragraph{Per-dataset dictionary size.}  Each of the four benchmarks has its own schema map $\mathcal{S}_{\mathrm{dataset}}$ and phrase dictionary $\mathcal{D}_{\mathrm{dataset}}$, authored once at onboarding from the item-side metadata vocabulary of that benchmark.  Coverage (the fraction of items for which a given channel of $\phi_i$ is non-empty after R1--R4) is high on the three structured domains (Yelp, MovieTV, Books) where item metadata carries controlled categorical tags, and noticeably lower on Goodreads (especially on the \texttt{context} and \texttt{audience} channels) because item metadata there is dominated by free-text summaries rather than categorical tags.  This is the same structural reason the Dialogue Gain on Goodreads is the weakest of the four benchmarks (\S\ref{sec:experiments}, Figure~\ref{tab:ladder_cross}).

\section{Additional Results}
\label{app:rebuttal_results}

All results below use $1{,}000$-user subsamples and the same protocol as \S\ref{sec:experiments}.

\paragraph{Other LLM backbones.}
Under identical code and candidate pools, \ours{} outperforms MemRec on every dataset with three further backbones (Table~\ref{tab:backbones}), including the open-weight qwen3-30b.

\begin{table}[h]
\centering
\small
\caption{Hit@1 of MemRec / \ours{} with other backbones.}
\label{tab:backbones}
\begin{tabular}{l|ccc}
\toprule
Dataset & deepseek-v3.2 & gemini-3-flash & qwen3-30b \\
\midrule
Yelp      & .510 / \textbf{.586} & .424 / \textbf{.471} & .479 / \textbf{.647} \\
MovieTV   & .545 / \textbf{.651} & .474 / \textbf{.527} & .481 / \textbf{.590} \\
Books     & .393 / \textbf{.584} & .578 / \textbf{.596} & .507 / \textbf{.596} \\
Goodreads & .347 / \textbf{.418} & .283 / \textbf{.392} & .318 / \textbf{.389} \\
\bottomrule
\end{tabular}
\end{table}

\paragraph{Where the profile is consumed.}
All methods rank the same fixed candidate pool (the ground truth plus 9 sampled negatives), so gains come from ranking the ground truth higher within the pool.  With only the consumption locus changed, retrieval-side gating improves Hit@1 over \textsc{Memory} by $+5.5$ / $+2.9$ pp on Yelp / MovieTV, more than reranker-side injection (Table~\ref{tab:ablation}).

\paragraph{Sensitivity.}
With uniform channel weights the gain over \textsc{Memory} remains positive on all four datasets but is smallest; the default and a preference-heavy setting differ by at most $0.5$ pp.  Stronger or weaker homophily floors also keep a positive gain, and the default floors lie near the top of this band.  Across candidate-pool sizes $|C|\in\{5,10,20\}$, \ours{} stays above both \textsc{Memory} and \textsc{Dialogue} on Yelp.

\paragraph{Scope.}
Our evaluation follows the accuracy-oriented InstructRec protocol.  The floor $\alpha$ of the homophily gate offers a simple way to soften the gate where exploration matters, for example for users with short histories, and beyond-accuracy measures such as novelty and diversity are a natural extension of this evaluation.  Profiles are built only from observed interactions, but may inherit biases of the LLM backbone.

\paragraph{Efficiency.}
On 1{,}000 Yelp queries with gpt-4o-mini, ranking a query takes two LLM calls (Stage-R and rerank), with a memory write when feedback is observed, as in MemRec; mean end-to-end latency is $10.81$\,s ($p_{95}=13.28$\,s), and pruning with the homophily gate takes about $0.6$\,ms.  Item profiles are rule-extracted offline and user profiles are built once per user.

\end{document}